\documentclass [12pt,eqsecnum,amsfonts,amssymb,aps]{revtex4-2}

\input epsf

\usepackage[usenames]{color}
\usepackage{graphicx}
\usepackage{bm}
\usepackage{rotating}

\newcommand{\beq}{\begin{equation}}
\newcommand{\eeq}{\end{equation}}
\newcommand{\beqs}{\begin{eqnarray}}
\newcommand{\eeqs}{\end{eqnarray}}

\begin{document}

\baselineskip 6.0mm

\title{Potts Partition Function Zeros and Ground State Entropy on Hanoi Graphs}

\author{Shu-Chiuan$^a$ Chang and Robert Shrock$^b$}

\affiliation{(a) \ Department of Physics \\
National Cheng Kung University, Tainan 70101, Taiwan}

\affiliation{(b) \ C. N. Yang Institute for Theoretical Physics and \\
  Department of Physics and Astronomy \\
  Stony Brook University, Stony Brook, NY 11794}

\begin{abstract}

  We study properties of the Potts model partition function $Z(H_m,q,v)$ on
  $m$'th iterates of Hanoi graphs, $H_m$, and use the results to draw
  inferences about the $m \to \infty$ limit that yields a self-similar Hanoi
  fractal, $H_\infty$.  We also calculate the chromatic polynomials
  $P(H_m,q)=Z(H_m,q,-1)$. From calculations of the configurational degeneracy,
  per vertex, of the zero-temperature Potts antiferromagnet on $H_m$, denoted
  $W(H_m,q)$, estimates of $W(H_\infty,q)$, are given for $q=3$ and $q=4$ and
  compared with known values on other lattices.  We compute the zeros of
  $Z(H_m,q,v)$ in the complex $q$ plane for various values of the
  temperature-dependent variable $v=y-1$ and in the complex $y$ plane for
  various values of $q$.  These are consistent with accumulating to form loci
  denoted ${\cal B}_q(v)$ and ${\cal B}_v(q)$, or equivalently, ${\cal
    B}_y(q)$, in the $m \to \infty$ limit.  Our results motivate the inference
  that the maximal point at which ${\cal B}_q(-1)$ crosses the real $q$ axis,
  denoted $q_c$, has the value $q_c=(1/2)(3+\sqrt{5} \, )$ and correspondingly,
  if $q=q_c$, then ${\cal B}_y(q_c)$ crosses the real $y$ axis at $y=0$, i.e.,
  the Potts antiferromagnet on $H_\infty$ with $q=(1/2)(3+\sqrt{5} \, )$ has a
  $T=0$ critical point.  Finally, we analyze the partition function zeros in
  the $y$ plane for $q \gg 1$ and show that these accumulate approximately
  along parts of the sides of an equilateral triangular with apex points that
  scale like $y \sim q^{2/3}$ and $y \sim q^{2/3} e^{\pm 2\pi i/3}$. Some
  comparisons are presented of these findings for Hanoi graphs with
  corresponding results on $m$'th iterates of Sierpinski gasket graphs and the
  $m \to \infty$ limit yielding the Sierpinski gasket fractal.
  
\end{abstract}

\maketitle

\newpage
\pagestyle{plain}
\pagenumbering{arabic}


\section{Introduction}
\label{intro_section}

Studies of iteratively defined hierarchical graphs $G_m$ with the property that
the limiting graph $G_\infty$ is a self-similar fractal have produced many
interesting results in physics and mathematics (some reviews include
\cite{mandelbrot}-\cite{falconer}).  There have been a number of studies of spin
models and critical phenomena on fractals, e.g., \cite{dhar}-\cite{irht}.  A
spin model of particular interest is the Potts model \cite{potts,fk,wurev}.  On
a lattice, or, more generally, on a graph $G$, in thermal equilibrium at
temperature $T$, the partition function for the Potts model is
\beq
Z= \sum_{\{\sigma_i\}}e^{-\beta {\cal H}} \ , 
\label{z}
\eeq
where $\beta = 1/(k_BT)$, $k_B$ is the Boltzmann constant, and the
Hamiltonian is
\beq
{\cal H} = -J\sum_{e_{ij}}\delta_{\sigma_i \sigma_j} \ ,
\label{ham}
\eeq
where $J$ is the spin-spin
interaction constant, $i$ and $j$ denote vertices (= sites) in $G$, $e_{ij}$ is
the edge (= bond) connecting them, and $\sigma_i$ are classical spins taking on
values in the set $\{1,...,q\}$. We use the notation
\beq
K = \beta J \ , \quad  y = e^K \ , \quad v = y -1 \ . 
\label{kv}
\eeq
We denote the partition function of the Potts model on a graph $G$ as
$Z(G,q,v)$. This function is equivalent to an important graph-theoretic
function, the Tutte polynomial, as will be reviewed in Section
\ref{background_section}. For the Potts antiferromagnet (PAF), $J < 0$ so that,
as $T \to 0$, $K \to -\infty$; hence, in this limit (where $v \to -1$), the
only contributions to the PAF partition function are from spin configurations
in which adjacent spins have different values.  The resultant $T=0$ PAF
partition function is therefore precisely the chromatic polynomial $P(G,q)$ of
the graph $G$, which counts the number of ways of assigning $q$ colors to the
vertices of $G$, subject to the condition that no two adjacent vertices have
the same color. An important feature of the antiferromagnetic Potts model is
that for sufficiently large $q$ on a given graph $G$ with finite maximal vertex
degree, it has a nonzero entropy per site at zero temperature, $S_0=k_B \ln W$,
where $W$ denotes the ground state degeneracy per site. This is important as an
exception to the third ``law'' of thermodynamics, that the entropy per site
vanishes at zero temperature. A physical example of this phenomenon is the
residual entropy of ice \cite{pauling}-\cite{lieb}.

A standard way to define a fractal is to start with some initial graph $G_0$
and then apply a procedure to construct a related graph with more vertices and
edges, forming the $m=1$ iterate, $G_1$, so forth with $G_2$, etc. By
continuing this process in an iterative manner, one produces a graph $G_m$, the
$m$'th iterate in the given hierarchical family.  In the cases of interest
here, in the limit $m \to \infty$, the resultant object, denoted $G_\infty$, is
self-similar, often with a non-integer Hausdorff dimension, whence the term
``fractal''.  Two graph iterates whose $m \to \infty$ limits yield fractals are
the $m$'th iterates of the Sierpinski gasket graph, $S_m$, and of the Hanoi
graph, $H_m$.  Sierpinski gasket iterates were studied in some of the earliest
papers on spin models on fractals \cite{gamprl}-\cite{gam_sg}.
Mathematical studies of Hanoi graphs include, e.g.,
\cite{jakovac}-\cite{hinz_survey}.  The first few iterates of Hanoi graphs are
shown below, using a common labelling convention in which the initial graph
is labelled $m=0$. (Some authors use a different labelling convention in which
the initial graph is denoted $m'=1$, so $m'=m+1$.)  

\bigskip
\bigskip

\unitlength 1mm
\begin{picture}(78,35)
\put(0,0){\line(1,0){6}}
\put(0,0){\line(3,5){3}}
\put(6,0){\line(-3,5){3}}
\put(3,-4){\makebox(0,0){$H_0$}}
\put(12,0){\line(1,0){18}}
\put(12,0){\line(3,5){9}}
\put(30,0){\line(-3,5){9}}
\put(18,10){\line(1,0){6}}
\put(24,0){\line(3,5){3}}
\put(18,0){\line(-3,5){3}}
\put(21,-4){\makebox(0,0){$H_1$}}
\put(36,0){\line(1,0){42}}
\put(36,0){\line(3,5){21}}
\put(78,0){\line(-3,5){21}}
\put(48,20){\line(1,0){18}}
\put(60,0){\line(3,5){9}}
\put(54,0){\line(-3,5){9}}
\multiput(42,10)(24,0){2}{\line(1,0){6}}
\multiput(48,0)(24,0){2}{\line(3,5){3}}
\multiput(42,0)(24,0){2}{\line(-3,5){3}}
\put(54,30){\line(1,0){6}}
\put(60,20){\line(3,5){3}}
\put(54,20){\line(-3,5){3}}
\put(57,-4){\makebox(0,0){$H_2$}}
\end{picture}

\bigskip
\bigskip

\begin{figure}[htbp]
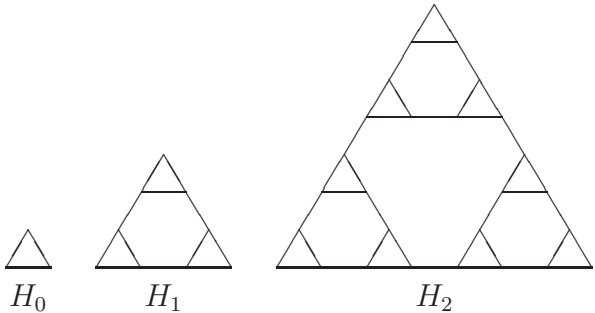

\caption{Initial Hanoi graph $H_0$ and first two iterates, $H_1$ and $H_2$}. 
\label{hanoi_figure}
\end{figure}

\bigskip

In this paper we study properties of the Potts model partition function
$Z(H_m,q,v)$ of $m$'th iterates of Hanoi graphs, $H_m$. We make use of an
iterative procedure for calculating the equivalent Tutte polynomial derived by
Donno and Iacono in Ref.  \cite{donno_iacono}. We also calculate the chromatic
polynomials $P(H_m,q)=Z(H_m,q,-1)$. From calculations of the configurational
degeneracy per vertex for the $q$-state Potts antiferromagnet, denoted
$W(H_m,q)$, for $q=3$ and $q=4$, for a large range of $m$, we extrapolate to $m
\to \infty$ to obtain estimates of $W(H_\infty,3)$ and $W(H_\infty,4)$, and
compare these with known values on other lattices. We calculate zeros of
$Z(H_m,q,v)$ in the $q$ plane for various values of $v$ and zeros of
$Z(H_m,q,v)$ in the $v$ plane for various values of $q$, including results up
to $m=4$. Our focus is primarily on the chromatic zeros, i.e., the zeros in the
complex $q$ plane for $v=-1$, corresponding to the zero-temperature Potts
antiferromagnet.  From our calculations, we are able to draw plausible
inferences concerning properties of the respective accumulation sets of zeros
in these respective planes in the limit $m \to \infty$, denoted ${\cal B}_q(v)$
and ${\cal B}_v(q)$, or equivalently, ${\cal B}_y(q)$, where $y=v+1$. As will
be discussed in detail below, our results motivate the inference that for
$v=-1$ (the zero-temperature Potts antiferromagnet), the maximal point at which
${\cal B}_q(v)$ crosses the real-$q$ axis, denoted $q_c$, has the value
$q_c=(1/2)(3+\sqrt{5} \, )$, and correspondingly, if $q=q_c$, then ${\cal
  B}_v(q_c)$ crosses the real $v$ axis at $v=-1$. Results are also given for
zeros of $Z(H_m,q,v)$ in the $q$ plane for the finite-temperature Potts
antiferromagnet and ferromagnet.  Furthermore, we analyze the partition
function zeros in the $y$ plane for several values of $q$ and for $q \gg 1$ and
determine the general behavior of the large-$q$ zeros. In previous work, we
have calculated partition function zeros on $m$'th iterates of the Sierpinski
gasket graph, and at appropriate points we will make comparisons with our
eaerlier results on Sierpinski gasket graphs. In addition to studies of spin
models on $m$'th iterates of hierarchical graphs, calculations of several
graph-theoretic quantities, such as the number of spanning trees, spanning
forests, connected spanning subgraphs, etc. have been computed on Sierpinski
and Hanoi graphs; some of these computations were presented in
\cite{sts}-\cite{dmth}.


\section{ Some Background}
\label{background_section}

In this section we briefly review some relevant background on graph theory, 
the Potts model partition function, Tutte polynomials, and Hanoi graphs. 
In general, a graph $G=(V,E)$ is defined by its set
of vertices (sites), $V$, and its set of edges (bonds), $E$.  We denote the
number of vertices in $G$ as $n=n(G)=|V|$ and the number of edges in $G$ as
$e(G)=|E|$. The degree $\Delta_{v_i}$ of a vertex $v_i \in V$ is defined as the
number of edges that connect to this vertex.  The number of connected
components of a graph, each of which is disjoint from the others, is denoted
$k(G)$.  The girth $g$ of a graph $G$ is the number of edges in a
minimal-length closed circuit in $G$. The cyclomatic number of $G$, i.e., the
number of linearly independent circuits in $G$, denoted $c(G)$,
and satisfies the relation $c(G)=e(G)+k(G)-n(G)$.  A graph $G'=(V,E')$ is a
spanning subgraph of a graph $G=(V,E)$ if it has the same vertex set and its
edge set is a subset of the edge set of $G$, i.e., $E' \subseteq E$
\cite{bbook,boll}.


\subsection{Potts Model Partition Function and Tutte Polynomial} 
\label{potts_tutte_section}

As defined in Eqs. (\ref{z}) and (\ref{ham}), the number of states of a given
classical spin in the Potts model is a positive integer, $q$. An important
generalization of this starts with an expression for the Potts model partition
function $Z(G,q,v)$ on a graph $G$ as a sum of contributions from spanning
subgraphs $G' \subseteq G$ \cite{fk}, which reads 
\beq
Z(G,q,v) = \sum_{G' \subseteq G} q^{k(G')}v^{e(G')} \ , 
\label{cluster}
\eeq
where $k(G')$ and $e(G')$ denote the number of connected components
and edges of $G'$.  Since $k(G') \ge 1$ and $e(G') \ge 0$, $Z(G,q,v)$
is a polynomial in $q$ and $v$ (of degree $n(G)$ in $q$ and of degree
$e(G)$ in $v$). The physical range of $v$ in the Potts ferromagnet
(FM, $J > 0$) is $0 \le v \le \infty$, corresponding to $\infty \ge T \ge 0$,
while in the Potts antiferromagnet (AFM, $J < 0$), it is
$-1 \le v \le 0$ corresponding to $0 \le T \le \infty$. In the 
ferromagnetic case, Eq. (\ref{cluster}) allows one to extend   
the definition of $q$ from the positive integers ${\mathbb Z}_+$ to the
positive real numbers ${\mathbb R}_+$ 
while maintaining $Z(G,q,v) > 0$ and hence a Gibbs measure.
One can formally also consider this extension for the antiferromagnetic case,
although in this case, if $q$ is not a positive integer, it is not guaranteed
that $Z(G,q,v)$ is positive, since $v < 0$.  For 
the Potts antiferromagnet, since $J < 0$, as 
$T \to 0$, $K \to -\infty$ and $v \to -1$; as noted in the introduction, 
in this limit, the resultant $T=0$ partition function is the 
chromatic polynomial $P(G,q)$ of the graph $G$ \cite{bbook}-\cite{dkt}: 
\beq
Z(G,q,-1) = P(G,q) \ .
\label{zp}
\eeq
This coloring is called a proper $q$-coloring of (the vertices of)
$G$. The minimum value of $q$ for which one can perform a proper $q$-coloring
of a graph $G$ is the chromatic number of $G$, denoted $\chi(G)$.

On a given graph $G$, the ground state (i.e.,
zero-temperature) degeneracy per vertex of the Potts antiferromagnet is
\beq
W(G,q) = [P(G,q)]^{1/n(G)} \ . 
\label{wfinite}
\eeq
With physically relevant values of $q$, $P(G,q)$ is positive, and one uses the 
canonical real positive $n$'th root in this evaluation.
For the $n(G) \to \infty$ limit of a given
family of graphs $G$, we formally denote
\beq
\{ G \} = \lim_{n(G) \to \infty} G \ . 
\label{ginf}
\eeq
The ground state degeneracy per vertex of the Potts antiferromagnet in this
limit is
\beq
W(\{G\},q) = \lim_{n(G) \to \infty} [P(G,q)]^{1/n(G)} \ , 
\label{wdef}
\eeq
and the corresponding ground state entropy per site is 
\beq
S_0(\{ G \},q) = k_B \ln [W(\{G \},q)] \ . 
\label{s0}
\eeq
As discussed in \cite{w,a}, for certain values of $q$, denoted $q_s$, 
one must take account of the noncommutativity
\beq
\lim_{n(G) \to \infty} \lim_{q \to q_s} [P(G,q)]^{1/n(G)} \ne
\lim_{q \to q_s}\lim_{n(G) \to \infty} [P(G,q)]^{1/n(G)} \ .
\label{wnoncom}
\eeq
The special values of $q_s$ here include $q \in \{0,1,2\}$ since $P(H_m,q)$
vanishes at these values. Because the calculations of $W(H_m,q)$ that we 
present in this paper are for $q > 2$, either order of limits can be used
for these calculations. 

For a general graph $G$, the Potts model partition function is equivalent
to an important function in mathematical graph theory, the Tutte polynomial,
$T(G,x,y)$ \cite{tutte54}-\cite{bo}. The Tutte polynomial
can be expressed as a sum of contributions from spanning 
subgraphs $G' \subseteq G$ as
\beq
T(G,x,y) = \sum_{G' \subseteq G} (x-1)^{k(G')-k(G)}(y-1)^{c(G')} \ . 
\label{t}
\eeq
Since $k(G')-k(G) \ge 0$ and $c(G') \ge 0$, this is, indeed, a polynomial
in $x$ and $y$.  The equivalence between $Z(G,q,v)$ and $T(G,x,y)$ is
\beqs
Z(G,q,v) &=& (x-1)^{k(G)}(y-1)^{n(G)}T(G,x,y) \cr\cr
         &=& (q/v)^{k(G)}v^{n(G)}T(G,x,y) \ , 
\label{ztrel}
\eeqs
where 
\beq
x = 1 + \frac{q}{v}
\label{xqv}
\eeq
and
\beq
y = v+1 = e^K \ . 
\label{yqv}
\eeq
where $K$ was defined in Eq. (\ref{kv}). Note that 
\beq
q=(x-1)(y-1) \ .
\label{qxy}
\eeq
Special cases of the Tutte polynomial yield a number of important one-variable
polynomials, including the chromatic polynomial, flow polynomial, and
reliability polynomial.  In particular, the special case $x=1-q$, $y=0$ yields
the chromatic polynomial:
\beq
P(G,q) = (-q)^{k(G)}(-1)^{n(G)}T(G,1-q,0) \ . 
\label{chrompoly}
\eeq
%


\subsection{Hanoi Graphs}
\label{hanoi_section}

We recall some elementary properties of the $H_m$ hierarchical graphs.  The
initial graph $H_0$ is a triangle graph, $K_3 = C_3$, where $K_n$ and $C_n$ are
respectively, the complete graph and the circuit graph with $n$ vertices.  With
the commonly used labelling convention for $m$'th iterates that we follow,
the number of vertices in the $m$'th iterate Hanoi graph is
\beq
n(H_m) = 3^{m+1} \ ,  
\label{n_Hm}
\eeq
and the numbers of edges in $H_m$ is 
\beq
e(H_m) = \frac{3(3^{m+1}-1)}{2} \ .  
\label{e_Hm}  
\eeq
The cyclomatic number of $H_m$ is thus 
\beq
c(H_m) = \frac{3^{m+1}-1}{2} \ .
\label{c_Hm}
\eeq
The number of faces of $H_m$, denoted $N_F(H_m)$, is 
\beq
N_F(H_m) = \frac{3^{m+1}-1}{2} \ ,
\label{faces_Hm}
\eeq
which is equal to $c(H_m)$.  The number of triangular faces in
$H_m$, denoted $N_t(H_m)$, is
\beq
N_t(H_m)=3^m \ .
\label{triangles_Hm}
\eeq
Consequently, in the limit $m \to \infty$, the ratio of the number of
triangular faces to the total number of faces is
\beq
\lim_{m \to \infty} \frac{N_t(H_m)}{N_F(H_m)} = \frac{2}{3} \ .
\label{ntnf_Hinf}
\eeq

A $\Delta$-regular graph $G$ is defined as a graph with the property
that all of its vertices have the same degree, $\Delta$.  In a
$\Delta$-regular graph, one has the relation $\Delta = 2e(G)/n(G)$.
Even if a graph is not $\Delta$-regular, one can still define an
effective vertex degree in the $n(G) \to \infty$ limit as
\beq
\Delta_{\rm eff} = \lim_{n(G) \to \infty} \frac{2e(G)}{n(G)} \ . 
\label{delta_eff}
\eeq
In a Hanoi graph $H_m$, the three vertices forming the original $H_0$
triangle have vertex degree 2, while all of the other vertices have vertex
degree 3. Hence,
\beq
\Delta_{\rm eff}(H_\infty) = 3 \ .
\label{Delta_eff_Hinf}
\eeq

Since there are some similarities of Hanoi graphs and Sierpinski
graphs, it is useful to compare and contrast our new results
on zeros of $Z(H_m,q,v)$ with our previous results for the zeros of
$Z(S_m,q,v)$, where $S_m$ denotes the $m$'th iterate Sierpinski gasket graph.
For the reader's convenience, we list some relevant properties of
$m$'th iterates of Sierpinski gasket graphs in Appendix \ref{sg_appendix}. 


\section{Calculations and Results}
\label{results_section}

A nonlinear iterative procedure for calculating the Tutte polynomial
$T(H_m,x,y)$ of the $m$'th Hanoi graph iterate, $H_m$, in terms of
contributions from lower-$m$-order iterates, was given by Donno and Iacono in
\cite{donno_iacono} and is briefly described in Appendix
\ref{tcal_appendix}. (See also \cite{alvarez} for a different approach.)  Using
this nonlinear iterative method of Ref. \cite{donno_iacono}, we have calculated
$T(H_m,x,y)$ and the equivalent $Z(H_m,q,v)$ for $0 \le m \le 4$.  From Eq.
(\ref{zp}) or equivalently (\ref{chrompoly}), we have computed $P(H_m,q)$ for
these values of $m$. The results for the initial graph $H_0$ are elementary,
since $H_0=K_3=C_3$, so $T(H_0,x,y)=x+x^2+y$; $Z(H_0,q,v)=(q+v)^3+(q-1)v^3$;
and $P(H_0,q)= q(q-1)(q-2)$. We have used a different method to calculate the
values of chromatic polynomials and hence ground state entropy per vertex for
certain values of $q$ of particular interest, namely $q=B_5$ (see
Eq. (\ref{br})), $q=3$, and $q=4$; for these calculations; instead of having to
compute the full chromatic polynomial $P(H_m,q)$ for arbitrary $q$ and then
substitute a special value of $q$, we set $q$ equal to this value at the outset
in the iterative computation, which thus involves just integer arithmetic or
powers of elements of the algebraic number field ${\mathbb Q}[\sqrt{5} \,
]$. (Here the algebraic number field ${\mathbb Q}[\sqrt{t} \, ]$ is the field
of elements of the form $r+s\sqrt{t}$, where $r, \ s, \ t \in {\mathbb Q}$ and
$t$ is not a perfect square.)  These computations for fixed integers or
algebraic numbers can be carried to considerably higher values of $m$, as will
be discussed in Section \ref{w_section}.

We first discuss our results for the chromatic polynomials $P(H_m.q)$.  We find
the following general structural formula for $m \ge 1$ that describes the
$P(H_m,q)$ that we have calculated:
\beq
P(H_m,q)=q(q-1)(q-2)^{3m}Q_m(q) \quad {\rm for} \ m \ge 1 \ , 
\label{phm_form}
\eeq
where $Q_m(q)$ is a polynomial of degree
\beq
{\rm deg}(Q_m(q)) = 3^{m+1}-3m-2 \ .
\label{degsm}
\eeq
For $m=1$ we observe a simple factorization
\beqs
&& Q_1(q) = q^4-5q^3+10q^2-10q+5 \cr\cr
&=& \frac{1}{4}\Big [ 2q^2-(5+\sqrt{5} \, )q+(5+\sqrt{5} \, ) \Big ]
\Big [ 2q^2-(5-\sqrt{5} \, )q+(5-\sqrt{5} \, ) \Big ] \ . \cr\cr
&&
\label{Q1factored}
\eeqs
The polynomial $Q_2(q)$ that occurs in $P(H_2,q)$ is 
\beqs
&& Q_2(q) = q^{19}-26q^{18}+322q^{17}-2528q^{16}+14125q^{15}-59771q^{14}
\cr\cr
&+&198981q^{13}-534267q^{12}+1176423q^{11}-2147675q^{10}+3271840q^9
\cr\cr
&-&4170694q^8+4444555q^7-3940970q^6+2880770q^5-1709450q^4 \cr\cr
&+&803125q^3-286075q^2+70750q-9500 \ , 
\label{qq3}
\eeqs
In \cite{alvarez}, $P(H_m,q)$ was given for $m$ up to 2 (with our labelling
convention, which is equivalent $m'=3$ in the labelling convention of
Ref. \cite{alvarez}), and our results agree.  We have calculated $P(H_m,q)$ for
higher $m$ in a similar manner. From our results, we can observe several
interesting properties of $P(H_m,q)$ and its zeros, as well as drawing
plausible inferences for features of $P(H_\infty)$, from these results.

In addition to the structural property (\ref{phm_form}), one may
investigate factorizations of $P(H_m,q)$ (and thus $Q_m(q)$) for specific
values of $q$.  As an example, we take $q=3$ and $q=4$. We find that the values
of $P(H_m,q)$ do not, in general, have simple factorizations; for example,
\beq
P(H_3,3) = 2^{13} \cdot 3 \cdot (233) \cdot (32002057)
\label{phm3_q3_factorized}
\eeq
and
\beq
P(H_3,4) = 2^{24} \cdot 3 \cdot (34471) \cdot (67883) \cdot 
(12983) \cdot
(167772879347) \ . 
\label{phm3_q4_factorized}
\eeq
In contrast, in \cite{sg} we found that $P(S_m,3)=3! \ \forall \ m$ and that
$P(S_m,4)$ has simple factorizations as displayed in Eq. (\ref{psgm_q4})
below.


\section{Ground State Degeneracy of Potts Antiferromagnet on Hanoi Graphs}
\label{w_section}

From our calculations of the ground state degeneracy per vertex for the Potts
antiferromagnet on the Hanoi iterates, $H_m$, for a range of $m$, we can
extrapolate to $m \to \infty$ to obtain estimates of $W(H_\infty,q)$.  For
reference, we show values of $W(H_m,q)$ in Table \ref{whm_values} for our
inferred value of $q_c(H_\infty)=B_5$, (see Eq. (\ref{qc_hinf})) and for the
next two integral values of $q$, namely $q=3$ and $q=4$. (Here and below,
numbers in floating-point format are listed to the indicated number of
significant figures.) Because the integer arithmetic involved in the evaluation
of $P(H_m,3)$ and $P(H_m,4)$ is exact, while the
evaluation of $P(H_m,B_5)$, involving powers of the irrational quantity 
$q=B_5=(1/2)(3+\sqrt{5} \, )$, requires a
floating-point evaluation, we are able to obtain accurate
evaluations of $W(H_m,3)$ and $W(H_m,4)$ over a large range of $m$, 
namely $0 \le m \le 16$,
while we conservatively retain our evaluations of $W(H_m,q)$ for 
$q=B_5$ only up to $m=12$. 
Although we thus limit the listings in Table \ref{whm_values} to $0 \le m \le
12$, there is very little change in $W(H_m,q)$ going from $m=12$ to $m=16$
for $q=3$ and $q=4$, as is evident from 
$W(H_{16},3)=1.4887646$ versus $W(H_{12},3)=1.4887651$, and 
$W(H_{16},4)=2.4991903$ versus $W(H_{12},4)=2.4991909$. 
In Ref. \cite{alvarez}, values of $W(H_m,q)$
were given for $0 \le m \le 7$ (corresponding to $1 \le m' \le 8$ in the
labelling convention of \cite{alvarez}) and for some integral values of
$q$. For the range of $m$ where our values of $W(H_m,3)$ and $W(H_m,4)$ can be
compared with those in \cite{alvarez}, they agree, and ours extend to higher
$m$. We find that for the values of $m$ and $q$ for which we have performed
these calculations, $W(H_m,q)$ is a monotonically decreasing function of $m$
for fixed $q$. We consider the large-$m$ limit for two (integral) values of $q$
where comparison can be made with results for the zero-temperature $q$-state
Potts antiferromagnet on regular lattices, namely $q=3$ and $q=4$.
Extrapolating to the $m=\infty$ limit, we obtain $W(H_\infty,3) = 1.489(1)$ and
$W(H_\infty,4) = 2.499(1)$, where the estimated uncertainties are indicated in
parentheses.

\begin{table}[htbp]
  \caption{\footnotesize{ Values of the ground state degeneracy per vertex of
      the Potts antiferromagnet on the $m$'th Hanoi iterate, $H_m$, 
      for $q=(1/2)(3+\sqrt{5} \, )$, $q=3$, and $q=4$, denoted 
      $W(H_m,B_5)$, $W(H_m,3)$, and $W(H_m,4)$. Values are listed for 
      $0 \le m \le 12$}}
\begin{center}
\begin{tabular}{|c|c|c|c|} \hline\hline
$m$ & $W(H_m,B_5)$  & $W(H_m,3)$  & $W(H_m,4)$   \\ \hline
0   & 1.378241      & 1.817121      & 2.884499   \\ 
1   & 1.185301      & 1.592838      & 2.6219375  \\
2   & 1.123589      & 1.522681      & 2.539454   \\
3   & 1.103741      & 1.499985      & 2.512540   \\
4   & 1.097203      & 1.492495      & 2.503632   \\
5   & 1.095033      & 1.490007      & 2.500670   \\
6   & 1.094310      & 1.489179      & 2.499683   \\
7   & 1.094069      & 1.488903      & 2.499355   \\
8   & 1.093989      & 1.488811      & 2.499245   \\
9   & 1.093962      & 1.488780      & 2.4992085  \\
10  & 1.0939535     & 1.488770      & 2.499196   \\
11  & 1.093951      & 1.488766      & 2.499192   \\
12  & 1.093950      & 1.488765      & 2.499191   \\
\hline\hline
\end{tabular}
\end{center}
\label{whm_values}
\end{table}

It is of interest to compare these estimates with values of $W(\{G\},q)$ for $n
\to \infty$ limits of various families of graphs.  There have been many
calculations of $W(\{G\},q)$ and lower and upper bounds on this quantity for
various families of graphs, e.g., \cite{pauling}-\cite{lieb}, 
\cite{baxter70}-\cite{cjss} and later works (see
\cite{jemrev} for some references). Some calculations have been
carried out for hierarchical graphs $G_m$ leading to fractals in the $m \to
\infty$ limit in works including \cite{sg,chio_roeder,dhl}. As background for
comparisons, we remark on a general property of $W(\{G\},q)$. In the assignment
of colors to a given vertex of a graph, the constraint that this vertex must
have a color that is different from each adjacent vertex is more restrictive as
the number of adjacent vertices increases. If the graphs in a family are
$\Delta$-regular, then this number of adjacent vertices is given by
$\Delta$. Even if the graphs in a family are not $\Delta$-regular, since we
focus here on the $n(G) \to \infty$ limit, we may use $\Delta_{\rm eff}$ as a
measure of the number of adjacent vertices. Because the restriction on the
proper $q$-coloring of the vertices becomes more severe as $\Delta$ increases,
it follows that this reduces the ground state degeneracy, i.e., $W(\{G\},q)$ is
a decreasing function of $\Delta$ (or $\Delta_{\rm eff}$ if finite-$n$ graphs
are not $\Delta$-regular).  This monotonic dependence was shown in Fig. 5 of
\cite{w} for the honeycomb (hc), square (sq), and triangular (tri)
lattices. Since the coloring freedom is greater for larger $q$, one naturally
starts with $q \gg 1$ and moves to smaller values of $q$ in analyzing
$W(\{G\},q)$ for a particular limit $\{G\}$. The analytic form of $W(\{G\},q)$
is the same along the real-$q$ axis until $q$ decreases to
$q_c(\{G\})$. Therefore, if one compares $W(\{G\},q)$ for different $\{G\}$,
and, in particular, different regular lattices $\Lambda$, then the monotonicity
comparison can be made for the interval of $q$ larger than the largest
$q_c(\{G\})$ among the $\{G\}$ being compared. For these lattices, one knows
the integral values $q_c(tri)=4$ \cite{baxter86,baxter87}, $q_c(sq)=3$
\cite{lenard,lieb}, and, formally, $q_c(hc)=(1/2)(3+\sqrt{5} \, )$
\cite{p,p2}.  Thus, for the comparison, one takes the interval $q \ge 4$.  Over
this interval, the results in this Fig. 5 of \cite{w} show that for a
fixed $q$, $W(hc,q) > W(sq,q) > W(tri,q)$.  This set of inequalities is in
accord with the fact that $\Delta(hc) < \Delta(sq) < \Delta(tri)$.

\begin{table}[htbp]
  \caption{\footnotesize{ Values of $W(\{G\},3)$ and $W(\{G\},4)$ for
various $n(G) \to \infty$ limits of families of graphs and regular lattices 
$G$ that have $\Delta=3$ or $\Delta_{\rm eff}=3$. 
The quantity $g(G)$ is the girth.}}
\begin{center}
\begin{tabular}{|c|c|c|c|} \hline\hline
$\{G\}$                  & $g(G)$  & $W(\{G\},3)$  & $W(\{G\},4)$ \\ \hline
$H_\infty$               & 3       & 1.489(1)    & 2.499(1)     \\
$\Lambda_{488}$          & 4       & 1.686       & 2.622        \\
sq, $2_F \times \infty$  & 4       & 1.732       & 2.646        \\
hc                       & 6       & 1.660       & 2.604        \\
\hline\hline
\end{tabular}
\end{center}
\label{whm_comparison}
\end{table}

Since $\Delta_{\rm eff}=3$ for $H_\infty$, we first compare our estimates of
$W(H_\infty,3)$ and $W(H_\infty,4)$ with $W(\{G\},q)$ with $q=3, \ 4$ for $n(G)
\to \infty$ limits $\{G\}$ of several $\Delta$-regular families of graphs with
the same vertex degree, $\Delta=3$. This comparison is shown in Table
\ref{whm_comparison}. From the discussion in the previous paragraph, one
expects that, for a given $q$ in the interval larger than the largest
$q_c(\{G\})$ among the limits $\{G\}$ being compared, $W(H_\infty,q)$ should be
similar to $W(\{G\},q)$ for other $\{G\}$ with the same $\Delta$ or
$\Delta_{\rm eff}$, and the comparison in Table \ref{whm_comparison} is in
agreement with this expectation.  We comment on the entries in Table
\ref{whm_comparison} as follows.  The values of $W(hc,3)$ and $W(hc,4)$ are
from high-precision Monte Carlo simulations performed for \cite{ww} and
\cite{w3}. The actual Monte Carlo (MC) results given in \cite{ww,w3} (with
uncertainties in parentheses) are $W(hc,3) = 1.6600(5)$ and $W(hc,4) =
2.6038(7)$.  In Table \ref{whm_comparison} we also list a comparison with
$W(\Lambda_{488},q)$ with $q=3, \ 4$.  The actual values computed from MC
simulations for Ref. \cite{w3} are $W(\Lambda_{488},3) = 1.68575(60)$ and
$W(\Lambda_{488},4)=2.62226(75)$.  Here, $\Lambda_{488}$ is an Archimedean
lattice comprised of squares and octagons. We recall the definition of an
Archimedean lattice, as a uniform tiling of the plane by regular polygons in
which all vertices are equivalent.  Such a lattice is specified by the ordered
sequence of polygons that one traverses in making a complete circuit around a
vertex in a given (say counterclockwise) direction.  This is incorporated in
the mathematical notation for an Archimedean lattice, $(\prod_i p_i^{a_i})$,
where in the above circuit, the notation $p_i^{a_i}$ indicates that the regular
polygon with $p_i$ sides occurs contiguously $a_i$ times; it can also occur
noncontiguously. Thus, the $\Lambda_{488}$ lattice is the Archimedean lattice
such that when one makes a circuit in the local neighborhood of any vertex, one
traverses a square, and then two octagons.  In Table \ref{whm_comparison} we
also list values of $W(\{G\},q)$ with $q=3, \ 4$ for the infinite-length limit
of the strip of the square lattice with width $L_y=2$ vertices and free
transverse boundary conditions (which is independent of the longitudinal
boundary conditions), for which \cite{w}
\beq
W(sq,2_F \times \infty,q) = (q^2-3q+3)^{1/2} \ . 
\label{wlad}
\eeq
Hence, $W(sq,2_F \times \infty,3) = \sqrt{3} = 1.7320508..$ and $W(sq,2_F
\times \infty,4) = \sqrt{7} = 2.6457513..$ (In the case of periodic
longitudinal boundary conditions, this is a $\Delta$-regular graph, while in
the case of longitudinal boundary conditions, one uses $\Delta_{\rm eff}$, and
these are both equal to 3.)  We also note that for the Diamond Hierarchical
Lattice (DHL), with $\Delta_{\rm eff}=3$ (and girth 4), denoting the $m$'th
iterate as $D_m$, Ref. \cite{dhl} obtained
$W(D_\infty,3)=\sqrt{3}=1.7320508..$, which again is similar to the value of
$W(H_\infty,3)$.

We may also compare our inferred values of $W(H_m,3)$ and $W(H_m,4)$, as well
as our estimates of $W(H_\infty,3)$ and $W(H_\infty,4)$, 
with values for the Sierpinski gasket, with $\Delta_{\rm eff}=4$. For the 
$m$'th iterate $S_m$ of this family of hierarchical graphs, 
\beq
P(S_m,q=3) = 3!
\label{psgm_q3}
\eeq
so that
\beq
W(S_\infty,q=3)=1 \ . 
\label{wsinf_q3}
\eeq
One has 
\beq
P(S_0,q=4)=24 = 2^3 \cdot 3
\label{pgsm0_q4}
\eeq
and in \cite{sg}, for $m \ge 1$, we obtained 
\beq
P(S_m,q=4)=2^{3^m+3} \cdot 3^{3^{m-1}} \ , 
\label{psgm_q4}
\eeq
so that in the limit $m \to \infty$, the ground state (i.e., zero-temperature)
degeneracy per site for the Potts antiferromagnet on $S_\infty$ is
\beq
W(S_\infty,q=4) = 2^{2/3} \cdot 3^{2/9} = 2.026346 \ , 
\label{wsinf}
\eeq
in agreement with Ref. \cite{andrade93}, where $W(S_\infty,q=4)$ had been
obtained earlier.  The approach to this asymptotic limit is shown by the
specific values for $W(S_m,3)$ and $W(S_m,4)$ listed in Table
\ref{wsgm_values}. As is evident from Table \ref{wsgm_values}, these values
converge reasonably rapidly toward their respective $m \to \infty$ values.  Our
values of $W(H_m,3)$ and $W(H_m,4)$ in Table \ref{whm_values} also show
reasonably rapid convergence, which led to the quoted uncertainties in our
extrapolations to estimate the values of $W(H_\infty,3)$ and $W(H_\infty,4)$.

\begin{table}
  \caption{\footnotesize{ Values of the ground state degeneracy per vertex of
      the Potts antiferromagnet on $S_m$ for $q=3$ and $q=4$, denoted 
      $W(S_m,3)$ and $W(S_m,4)$, with $0 \le m \le 12$, for comparison with the values of $W(H_m,3)$ and $W(H_m,4)$ in Table \ref{whm_values}.}}
\begin{center}
\begin{tabular}{|c|c|c|} \hline\hline
$m$ & $W(S_m,3)$  & $W(S_m,4)$   \\ \hline
0   & 1.817121    & 2.884499   \\ 
1   & 1.348006    & 2.401874   \\
2   & 1.126878    & 2.1689435  \\
3   & 1.043584    & 2.076164   \\
4   & 1.014674    & 2.043221   \\
5   & 1.0049075   & 2.032002   \\
6   & 1.001638    & 2.028235   \\
7   & 1.000546    & 2.026976   \\
8   & 1.000182    & 2.026556   \\
9   & 1.000061    & 2.026416   \\
10  & 1.000020    & 2.026369   \\
11  & 1.000007    & 2.026354   \\
12  & 1.000002    & 2.026349   \\
$\infty$ & 1      & 2.026346   \\ 
\hline\hline
\end{tabular}
\end{center}
\label{wsgm_values}
\end{table}

In general, for the values of $m$ for which we have calculated $W(H_m,q)$ and
$W(S_m,q)$, we find the inequality for $m \ge 1$ 
\beq
W(H_m,q) > W(S_m,q) \quad {\rm if} \ m \ge 1 \ \ {\rm and} \ \ q \ge 3  \ ,
\label{whm_wsgm_inequality}
\eeq
The restriction to $m \ge 1$ is made here because $W(H_0,q) = W(S_0,q)$, as a
consequence of the fact that the initial graph for both $S_m$ and $H_m$
iterates is the same, namely a triangle: $S_0 = H_0 = K_3$. The inequality
(\ref{whm_wsgm_inequality}) reflects the above-mentioned property that, for a
fixed $q > q_c(\{G\})$, $W(\{ G \},q)$ is a monotonically decreasing function
of the vertex degree $\Delta$ or, where applicable, the effective vertex degree
$\Delta_{\rm eff}$. Here, the inequality can be understood since the fractal
$H_\infty$ has a smaller value of $\Delta_{\rm eff}$, namely 3, compared with
$S_\infty$, for which $\Delta_{\rm eff}=4$. Presuming that
(\ref{whm_wsgm_inequality}) holds for arbitrarily large $m$, it implies that in
the $m \to \infty$ limit,
\beq
W(H_\infty,q) \ge W(S_\infty,q) \quad {\rm for} \ q \ge 3 \ , 
\label{whinf_wsginf_inequality}
\eeq
and again, one expects this to be realized as a strict equality. 


\section{Chromatic Zeros of $H_m$}
\label{phmzeros_section}

\begin{figure}[htbp]
\begin{center}
\includegraphics[height=6cm]{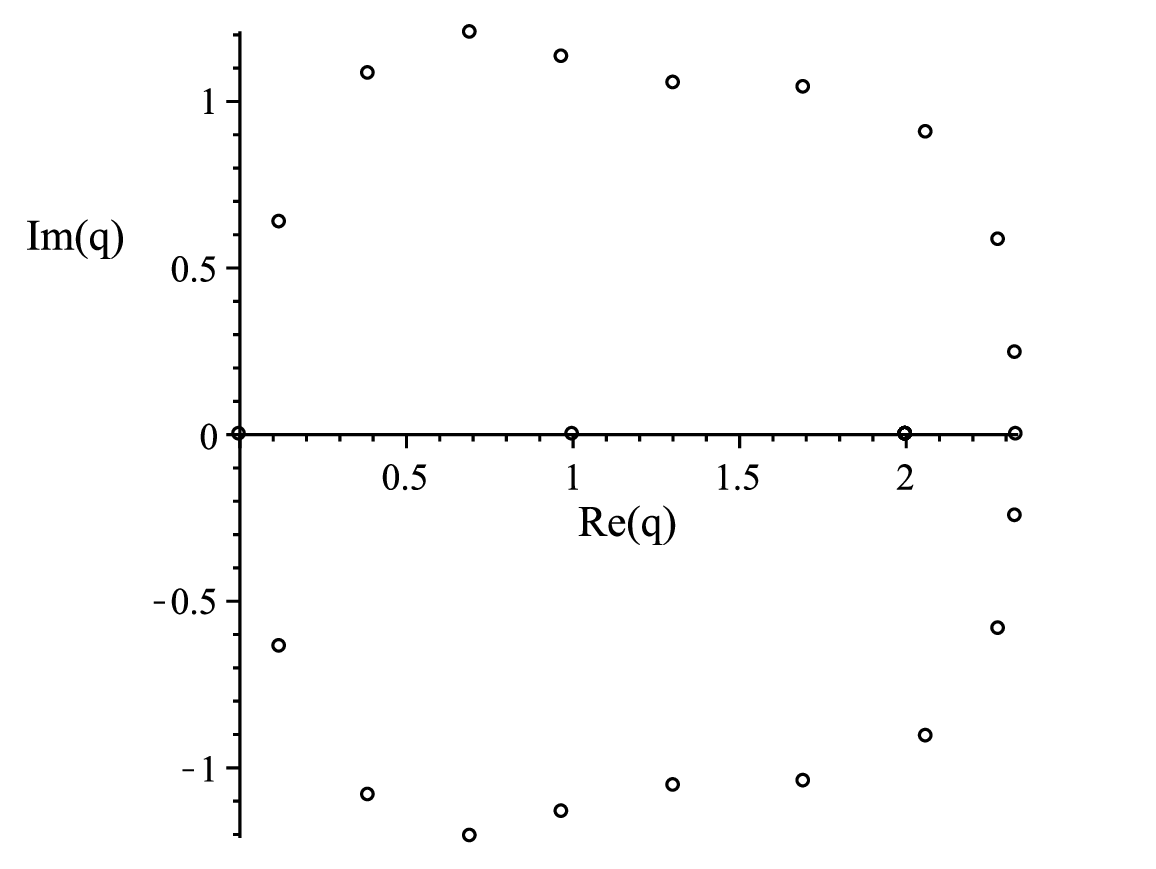}
\end{center}
\caption{\footnotesize{Zeros of the chromatic polynomial $P(H_2,q)$
in the $q$ plane.}}
\label{cphm2zeros_fig}
\end{figure}

\begin{figure}[htbp]
\begin{center}
\includegraphics[height=6cm]{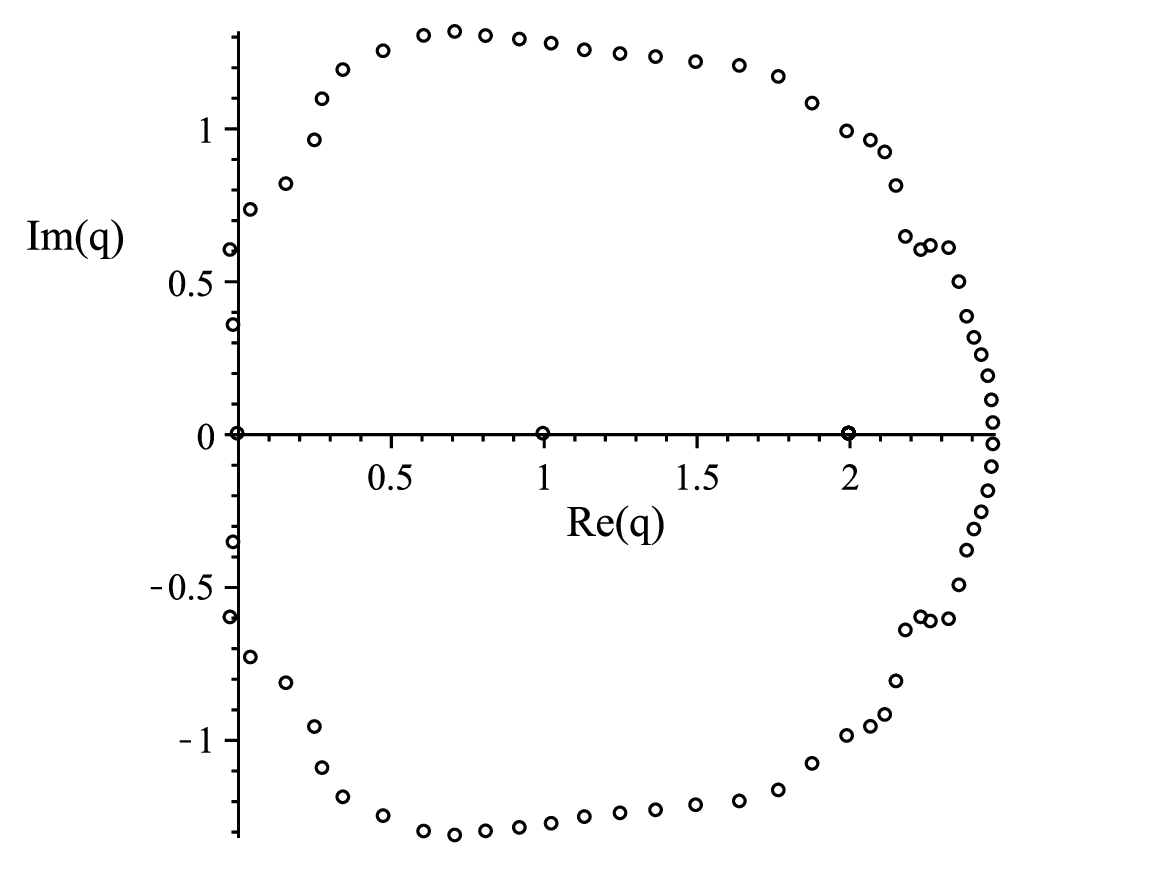}
\end{center}
\caption{\footnotesize{Zeros of the chromatic polynomial $P(H_3,q)$
in the $q$ plane.}}
\label{cphm3zeros_fig}
\end{figure}

\begin{figure}[htbp]
\begin{center}
\includegraphics[height=6cm]{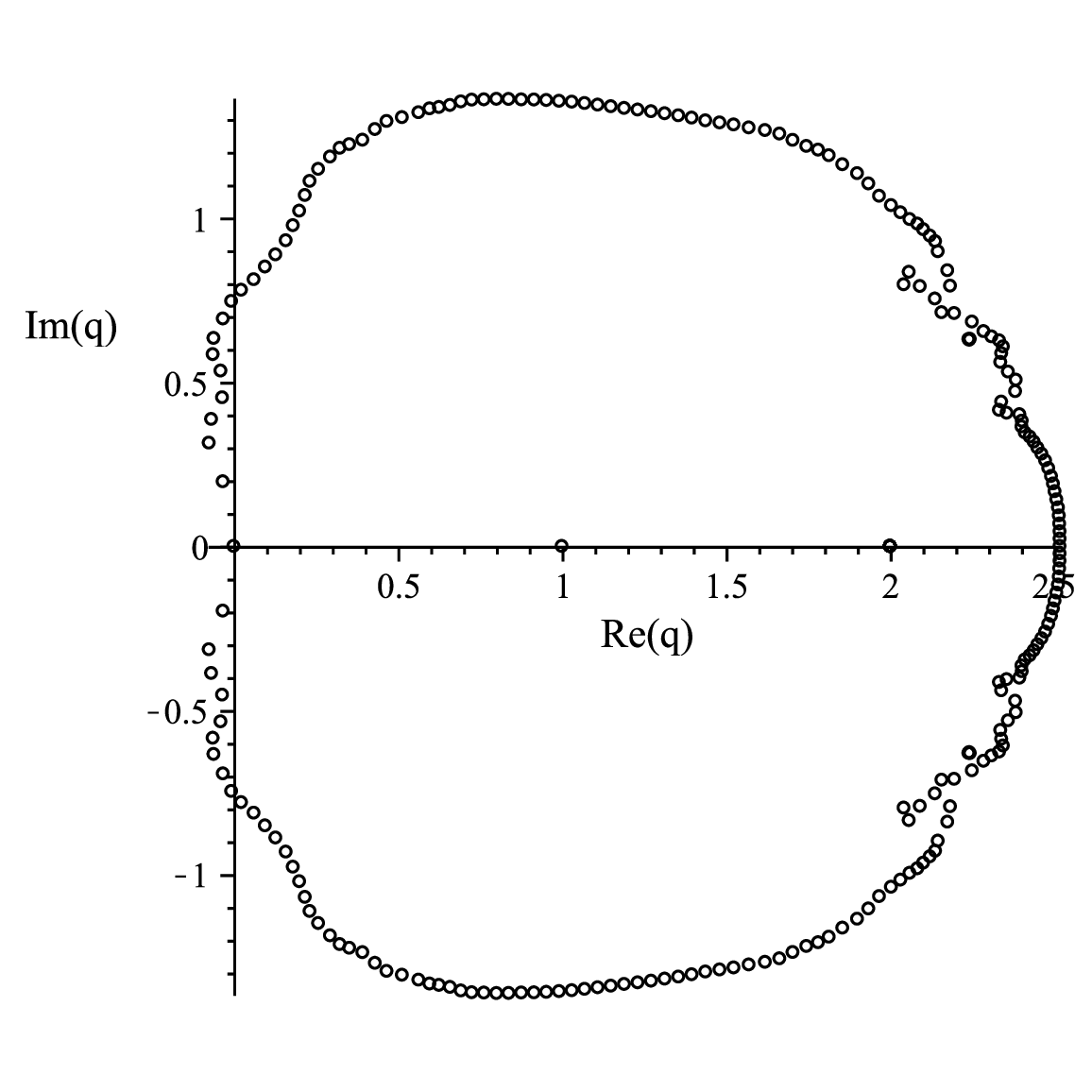}
\end{center}
\caption{\footnotesize{Zeros of the chromatic polynomial $P(H_4,q)$
in the $q$ plane.}}
\label{cphm4zeros_fig}
\end{figure}

In this section we study the zeros of $P(H_m,q)=Z(H_m,q,-1)$, i.e., the
chromatic zeros of $H_m$.  In Figs. \ref{cphm2zeros_fig}-\ref{cphm4zeros_fig}
we show plots of the zeros of $P(H_m,q)$ in the complex $q$ plane for $2 \le m
\le 4$.  As is evident, the complex zeros form a roughly oval shape centered
approximately at $q=1$.  It may be recalled that for an arbitrary graph, a
chromatic polynomial has the following zero-free regions on the real axis: (i)
$(-\infty,0)$, (ii) (0,1) \cite{woodall}, and (iii) $(1,\frac{32}{27})$
\cite{jackson,thomassen}.  The chromatic polynomials $P(H_m,q)$ always have
zeros at $q=0$, $q=1$, and $q=2$.  Most of the zeros have
positive real parts, although some zeros on the left-hand part of the locus
with nonzero imaginary parts have small negative real parts, i.e., lie in the
second and third quadrants. 

We denote the locus of zeros of $Z(H_m,q,v)$ in the limit $m \to
\infty$ (i) in the complex $q$ plane for a given $v$ as ${\cal
  B}_q(v)$ and (ii) in the complex $v$ plane for a given $q$ as ${\cal
  B}_v(q)$. Since we only discuss the chromatic zeros of $H_m$ in this
section, we will use the simpler notation for the asymptotic locus
${\cal B}_q(-1) = {\cal B}_q$, with $v=-1$ being understood
implicitly.  In a manner similar to our earlier study of zeros of the
partition function for Sierpinski gasket graphs in Ref. \cite{sg}, our
study here of chromatic zeros for a range of $m$, allows us to draw
some plausible inferences about the $m \to \infty$ limit. In
particular, we infer that ${\cal B}_q$ crosses the real-$q$ axis at
$q=0$ and at a maximal point which we denote $q_c$.  For finite $m$,
we denote the largest real (lr) zero of $P(H_m,q)$ as $q_{lr}$.  For
certain $m$, the zeros of $P(H_m,q)$ include a complex-conjugate pair
characterized by a small imaginary part and a real part that is larger
than any other real zeros or the real parts of other complex-conjugate
pairs of zeros.  We label the real parts of these complex-conjugate
pairs with the abbreviation (lrp), standing for ``largest real part''.

\begin{table}
  \caption{\footnotesize{ Values of largest real (lr) zero or largest
      real part (lrp) of the complex-conjugate pair of zeros of
      $P(H_m,q)$ for $0 \le m \le 4$, with extrapolated value in the
      $m \to \infty$ limit. For comparison with
      $B_5=(1/2)(3+\sqrt{5} \, )$, we list the respective fractional
      differences $FD = (B_5-q_{lr})/B_5$ or $FD=(B_5-q_{lrp})/B_5$ in
      the third column.}}
\begin{center}
\begin{tabular}{|c|c|c|} \hline\hline
$m$ &  $q_{lr}$ or $q_{lrp}$ & $FD$      \\ \hline
0   &  2 \ (lr)              & 0.236068  \\
1   &  2 \ (lr)              & 0.236068  \\
2   &  2.331734 \ (lr)       & 0.109357  \\
3   &  2.472039 \ (lrp)      & 0.055765  \\
4   &  2.517208 \ (lrp)      & 0.038512  \\
\hline\hline
\end{tabular}
\end{center}
\label{qm_values}
\end{table}

An extrapolation of these results leads to the inference that in the
$m \to \infty$ limit,
\beq
q_c(H_\infty) = \frac{3+\sqrt{5}}{2} = 2.618034 \ . 
\label{qc_hinf}
\eeq
The complex-conjugate pairs whose real parts are listed in Table \ref{qc_hinf} 
are
\beq
m=3: \quad q=2.472039 \pm 0.0349188i
\label{hm3_lrp_zero}
\eeq
and
\beq
m=4: \quad q=2.517208 \pm 0.0225300i \ .
\label{hm4_lrp_zero}
\eeq
Our use of the real parts of these complex-conjugate pairs to get
information relevant to $q_c$ is motivated by the fact that they
are quite close to the real axis; the ratios of the magnitude of the
respective imaginary parts divided by the real parts are
0.0141255 for $m=3$ and 0.008950 for $m=4$, decreasing as $m$ increases.
Our inferred value of $q_c(H_\infty)$ in Eq. (\ref{qc_hinf}) is
related to the well-known golden mean $\phi$
\beq
\phi = \frac{1+\sqrt{5}}{2} = 1.618034
\label{phi}
\eeq
satisfying $\phi^2 = \phi+1$. Namely,
\beq
q_c(H_\infty) = \phi+1 \ .
\label{qcgolden}
\eeq
It is also of interest to observe that our inferred value of $q_c(H_\infty)$ is
equal to a Tutte-Beraha number, namely
\beq
q_c(H_\infty) = B_5 \ ,
\label{qc_b5}
\eeq
where
\beq
B_r = 4 \cos^2\Big ( \frac{\pi}{r} \Big ) \ .
\label{br}
\eeq

Note that for any finite $m$, $B_5$ cannot be a chromatic zero of
$H_m$ or, indeed, any graph. We recall the elementary proof of
this. Assume that a quantity $q_0 = a+\sqrt{b}$ is a zero of a
chromatic polynomial $P(G,q)$ of a graph $G$ and assume that $a$ is
rational and $b> 0$ is rational but is not a perfect square, so $q_0$
is irrational.  Then since the coefficients of all terms in $P(G,q)$
are rational (indeed, are integers), it must be the case that the
algebraic conjugate, $a-\sqrt{b}$ is also a zero of $P(G,q)$ so that
the product $[q-(a+\sqrt{b} \, )][q-(a-\sqrt{b} \, )] = q^2-2aq +
(a^2-b)$ involves rational coefficients.  For $q=B_5$, this would
imply that $(1/2)(3-\sqrt{5} \, )=0.381966$ is a chromatic zero.
However, this is not possible because of the theorem \cite{woodall,dkt}
that the interval $q \in (0,1)$ is free of chromatic zeros.

Although the Hanoi fractal $H_\infty$ is a self-similar object rather
than a regular lattice, its effective vertex degree is $\Delta_{\rm
  eff}=3$.  A comparison with $q_c$ values for regular lattices with
various vertex degrees $\Delta$ or effective vertex degrees
$\Delta_{\rm eff}$ is thus of interest. The value that we infer for
$q_c(H_\infty)$ is equal to the value $q_c(hc)=(1/2)(3+\sqrt{5} \, )$
for the honeycomb lattice, which has $\Delta=3$ (see, e.g.,
\cite{p,p2}).  However, it should be cautioned that the infinite-$n$
limits of two families of $n$-vertex graphs with the same $\Delta$ may
have different values of $q_c$. Consider, for example, strips of the
square lattice with length $L_x$ vertices, transverse width $L_y$
vertices, and toroidal boundary conditions (i.e., periodic in the
longitudinal and transverse directions). These strips are all
$\Delta$-regular graphs for any $L_x$ and $L_y$. But in the
infinite-length limits $L_x \to \infty$ with fixed width $L_y$, the
$L_y=2$ strip of the square lattice has $q_c=2$ \cite{w}, the $L_y=3$
strip has $q_c=3$ \cite{tk}, and the $L_y=4$ strip has $q_c=2.7827657$
\cite{tor4}.  (The $L_y=2$ toroidal strip of the square lattice has
double transverse edges, and could be removed from this comparison by
stipulating that families of graphs to be used for the $q_c$
comparison in the respective $n \to \infty$ limits must not have
multiple edges, but this still leaves the $L_y=3$ and $L_y=4$ toroidal
square-lattice strips, which have no multiple edges, the same $\Delta$
value of 4, and different values of $q_c$.)  Furthermore, infinite-$n$
limits of $n$-vertex families of graphs with different $\Delta$ values
can have the same $q_c$. Some examples are provided by the comparison
of the $n \to \infty$ limit of the circuit graph $C_n$, which has
$\Delta=2$ and $q_c=2$, and the $L_x \to \infty$ limit of a $L_x
\times L_y$ strip of the square lattice with periodic longitudinal
($L_x$) and free tranverse $(L_y$) boundary conditions, which has
$\Delta=3$ and $q_c=2$. Another example is the $L_x \to \infty$ limit
of the homeomorphic expansion of the $L_y=2$ cyclic strip of the
square lattice with $s$ additional vertices added to each horizontal
edge, which has $\Delta_{\rm eff}$ ranging between 3 for $s=0$ and 2
for $s \gg 1$, but which has $q_c=2$ for all $s$ \cite{pg}.
Nonetheless, it is of interest that the $q_c(H_\infty)$ value that we
infer for the Hanoi fractal is equal to $q_c(hc)$, and $\Delta_{\rm
  eff}(H_\infty)=3$, equal to $\Delta(hc)=3$ for the honeycomb
lattice.

These patterns of chromatic zeros for the $m$'th iterate Hanoi graphs
are similar to the patterns that we found for the corresponding $m$'th
iterates of the Sierpinski graphs in \cite{sg}, with one difference
being our inferred $q_c(H_\infty)=(1/2)(3+\sqrt{5} \,)$, while our
inferred value of $q_c$ for $S_\infty$ in \cite{sg} was
$q_c(S_\infty)=3$.  A notable aspect of the patterns of zeros of these
chromatic polynomials $P(H_m,q)$ is the absence of complex-conjugate
pairs of zeros extending into the interior of the region inside of the
outer approximate envelope of zeros.  This is in contrast to our
results for chromatic zeros of Sierpinski iterates $S_m$ displayed in
Figs. 1 and 2 of Ref. \cite{sg}, where we showed the zeros of
$P(S_4,q)$ and $P(S_5,q)$. As is evident, e.g., in Fig. 2 of
Ref. \cite{sg}, there is a smaller oval-like ring of zeros crossing
the real axis at $q \simeq 2.6$ and $q \simeq 2.74$, and thus located
inward of the rightmost part of the outer envelope of zeros. The
pattern of zeros for $P(H_m,q)$ that we find also contrasts with the
results that we obtained with R. Roeder in \cite{dhl} for the
chromatic zeros of the $m$'th iterates of the Diamond Hierarchical
Lattices, $D_m$, including rigorous results for the accumulation locus
of zeros ${\cal B}_q$ in the $m \to \infty$ limit. To the left of the
crossing of ${\cal B}_q$ at $q_c(D_\infty)=3$, there are an infinite
number of zero-free regions and associated intervals on the real axis
separated from each other by crossings of ${\cal B}_q$, starting with
a zero-free region containing the real interval extending from $q=3$
downward to a crossing of ${\cal B}_q$ at
\beq
q = -\frac{1}{3}(1+3\sqrt{57} \, )^{1/3} + 
      \frac{8}{3(1+3\sqrt{57} \, )^{1/3}} +
\frac{5}{3} \simeq 1.638897 \ , 
\label{q1}
\eeq
then another zero-free region containing the real interval $q \in
(1.409700,1.638897)$, another crossing at 1.409700, and so forth.
This infinite series of progressively smaller zero-free regions,
associated zero-free real interval, and
crossing points approach the point $q=32/27$ from above.  (See Table
II in \cite{dhl} for a list of the first 10 crossing points.)  As is
evident from the explicit zeros shown in Fig. 3 of \cite{dhl}, there
is an indication of the first two of these infinitely many zero-free
intervals and crossing points from the zeros of $P(D_4,q)$. A similar
indication is visible in Fig. 1 of \cite{sg} showing the zeros of
$P(S_4,q)$. 

A triangulation is a graph all of whose faces are triangles.  Although
the initial Hanoi graph, $H_0$, is a triangle, the $H_m$ with $m \ge 1$
are not triangulations.  For example, $H_1$ contains three triangular
faces (c.f. Eq. (\ref{triangles_Hm})) and one 6-sided face; $H_2$
contains nine triangular faces, three 6-sided faces, and one 12-sided
face, and so forth for higher $m$.  As stated in
Eq. (\ref{ntnf_Hinf}), in the limit $m \to \infty$, 2/3 of the faces
are triangles. Given that triangles comprise a majority of the faces
in this limit, it is of interest to investigate how strong is the deviation
fron the Tutte upper bound for triangulation graphs. This 
bound is as follows \cite{tbound}: If $G$ is a (planar) triangulation, denoted
$G_t$, then
\beqs
&& |P(G_t,\phi+1)| \le (\phi-1)^{n(G_t)-5} \ , \quad i.e., \quad
   |P(G_t,B_5)| \le (B_5-2)^{n(G_t)-5} \ .
\cr\cr
&&
\label{tbound}
\eeqs
Since $B_5-2=0.6180... < 1$, this upper bound decreases exponentially rapidly
as a function of $n(G_t)$.  The Tutte upper bound (\ref{tbound}) is sharp,
since it is saturated for the simplest triangulation, namely a triangle graph,
$K_3$. For $K_3$, the bound is that $|P(K_3,B_5)| \le (B_5-2)^{-2} = B_5$, and
$P(K_3,B_5)=B_5$.  For graphs that are not triangulations, as well as graphs
for which a majority of the faces are triangles, it is of interest to
determine how close they come to saturating the Tutte upper bound
For this purpose, one defines the ratio \cite{tub,cpt}. 
\beq
r(G) = \frac{|P(G,B_5)|}{(B_5-2)^{n(G)-5}} \ .
\label{rtub}
\eeq
The bound (\ref{tbound}) is thus the statement that if $G$ is a triangulation,
$G_t$, then $r(G_t) \le 1$.  We find that for $m \ge 2$ where $H_m$ is not
a triangulation, the ratio $r(H_m)$ is considerably larger than 1.


\section{Zeros of $Z(H_m,q,v)$ in the $q$ Plane at Finite Temperature}
\label{zhmzeros_section}

\begin{figure}[htbp]
\begin{center}
\includegraphics[height=6cm]{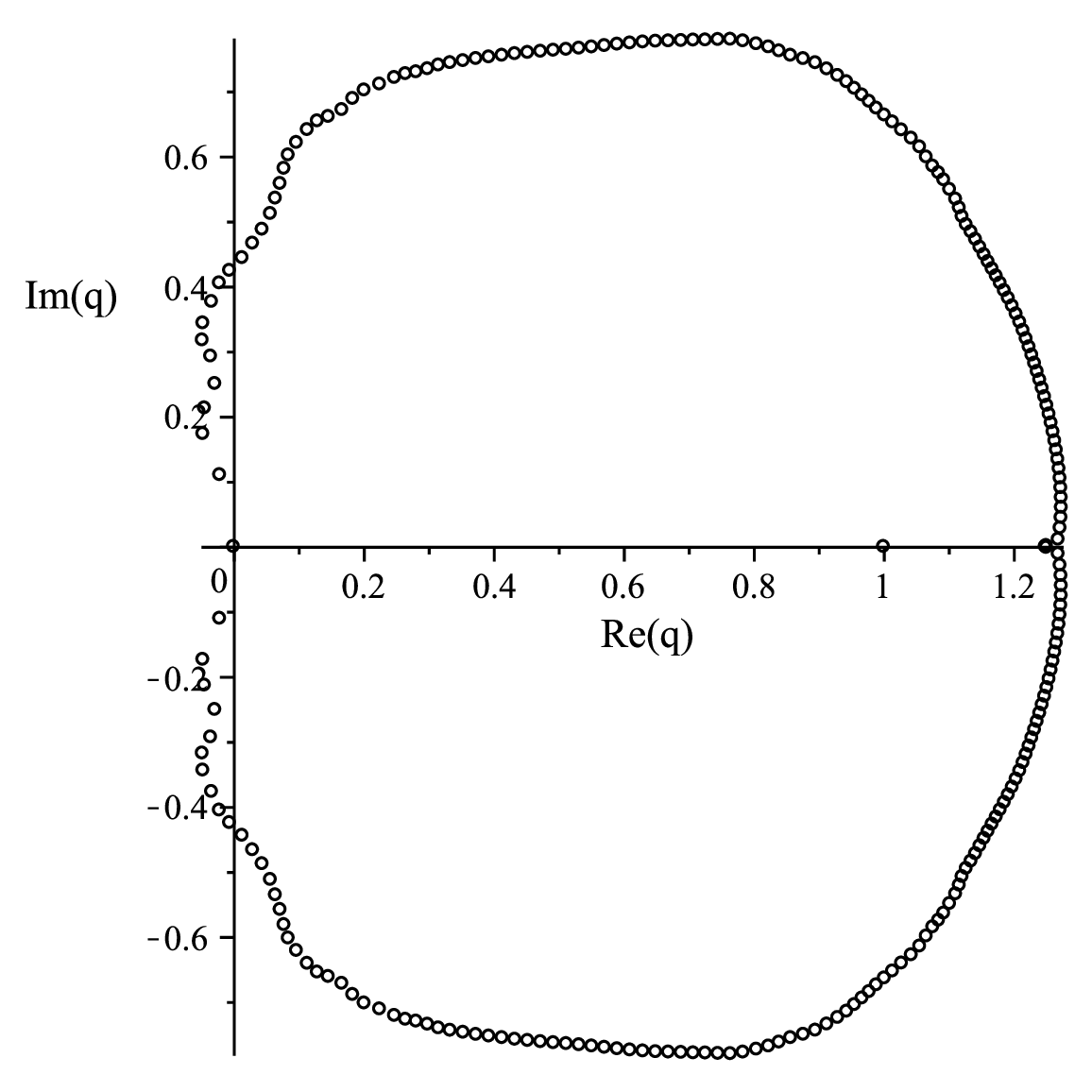}
\end{center}
\caption{\footnotesize{Zeros of the Potts partition function $Z(H_4,q,v)$
    in the $q$ plane for $v=-0.5$.}}
\label{hm4qplotvm0p5_fig}
\end{figure}

As stated in Eq. (\ref{zp}), for an arbitrary graph $G$, the chromatic
polynomial $P(G,q)$ is equal to the partition function of the zero-temperature
Potts antiferromagnet, $Z(G,q,v=-1)$.  As the temperature $T$ increases from 0
to $\infty$ for the Potts antiferromagnet, $v$ increases from $-1$ to $0^-$. In
Fig. \ref{hm4qplotvm0p5_fig} we show a plot of zeros of $Z(H_4,q,v)$ in the $q$
plane for $v=-0.5$, an illustrative finite-temperature value of $v$ for the
Potts antiferromagnet.  The pattern of zeros is smoother than for $v=-1$, and
it is contracting toward $q=0$.  As $v \to 0^-$, the zeros all move in toward
$q=0$, in accord with the general property that for an arbitrary graph $G$, if
$v=0$, then $Z(G,q,v=0)=q^{n(G)}$, so that all of the zeros occur at $q=0$.

\begin{figure}[htbp]
\begin{center}
\includegraphics[height=6cm]{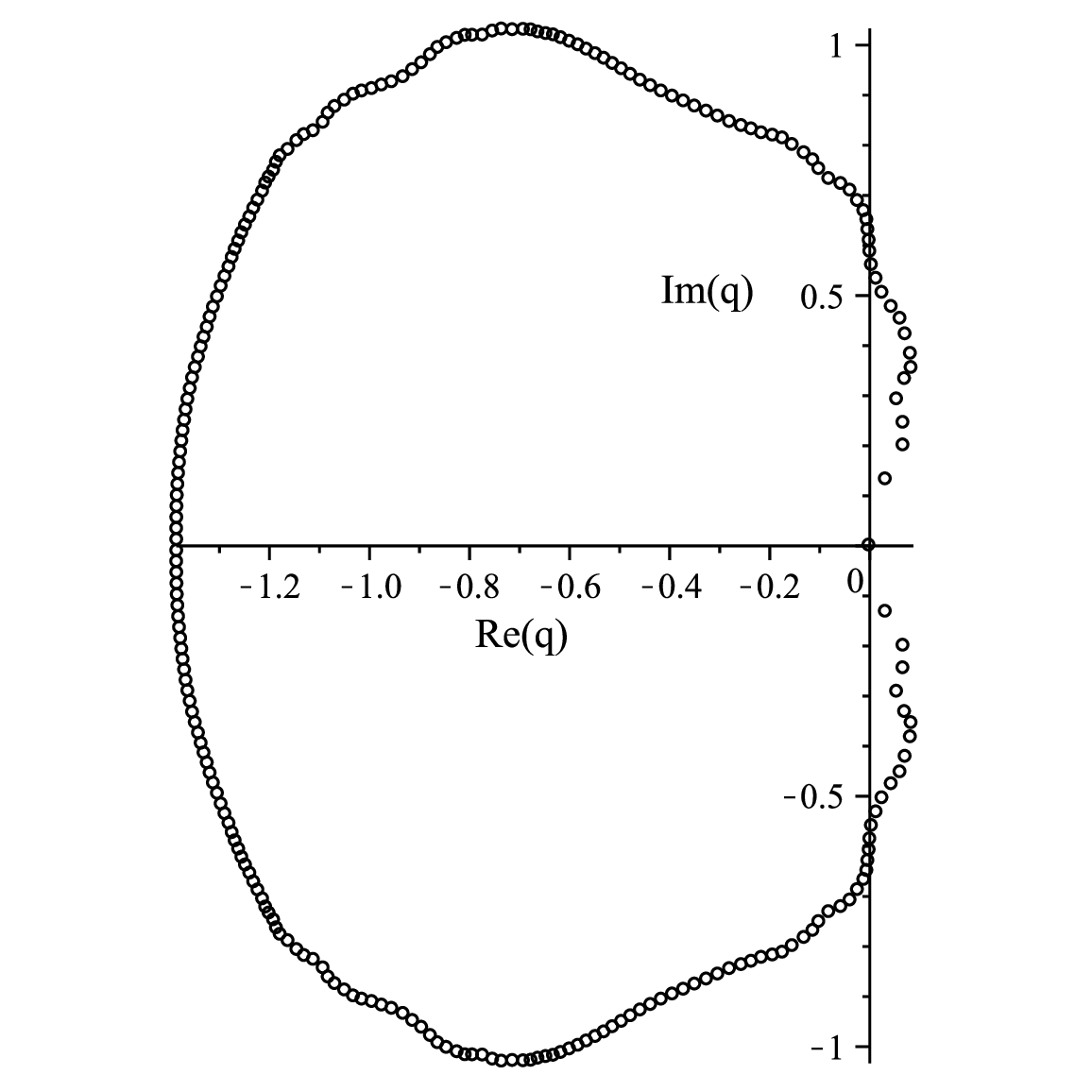}
\end{center}
\caption{\footnotesize{Zeros of the Potts partition function $Z(H_4,q,v)$
in the $q$ plane for $v=0.5$.}}
\label{zhm4zeros_v0p5_fig}
\end{figure}

As $T$ decreases from $\infty$ through finite values for the Potts ferromagnet,
$v$ increases from 0 through positive values.  In Fig. \ref{zhm4zeros_v0p5_fig}
we show a plot of the zeros of $Z(H_4,q,v)$ in the $q$ plane for a
representative finite-temperature value of $v$ for the Potts ferromagnet,
namely $v=0.5$. In this case, the zeros again exhibit a roughly oval form, and,
aside from the zero at $q=0$, most of them have negative real parts (some zeros
with nonzero imaginary parts have small positive real parts).


\section{Zeros of $Z(H_m,q,y)$ in the $y$ Plane}
\label{zhm_zeros_section}

It is also of interest to investigate the zeros of $Z(H_m,q,v)$ in the complex
plane of the temperature-dependent Boltzmann variable $v$, and we have
carried out this study.  For convenience, we will plot these zeros in the plane
of the variable $y=v+1=e^K$ and will use the notation
\beq
Z(G,q,y) \equiv Z(G,q,v)_{v=y-1} \ .
\label{zvy}
\eeq
As a historical note, it may be recalled that zeros of partition
functions of spin models on regular lattices have long been of
interest, dating from the pioneering analyses by Lee and Yang
\cite{yl,ly} of zeros of the Ising model in the plane of a Boltzmann
variable $z=e^{\beta H}$, where $H$ is an external magnetic field, and
an analysis by Fisher of zeros of the partition function of the Ising
model (in zero external magnetic field) in a temperature-dependent
Boltzmann variable \cite{fisher}. It was shown in early work
\cite{gamprl}-\cite{gam_ramified} that a necessary condition for a
discrete spin model on the $m \to \infty$ limit of a hierarchical
family of graphs $G_m$ to have an order-disorder phase transition at
finite temperature is that in this limit, the resultant fractal has
vertices with an infinite ramification number, $R$.  Here, as defined
in \cite{gamprl}-\cite{gam_ramified}, the ramification number $R$ of a
given vertex in a hierarchical graph $G_m$ is the number of edges that
must be cut to isolate the vertex from the rest of the
graph. (Different vertices may have different ramification numbers.)
It was noted in \cite{gamprl,gam_sg} that the $R$ numbers for vertices
in the Sierpinski gasket fractal are finite, and therefore the Potts
model (with either sign of $J$) does not have a physical
finite-temperature phase transition on $S_\infty$.  The same property
holds for Hanoi graphs, and hence the Potts model with either sign of
$J$ does not have a finite-temperature order-disorder transition on
the $H_\infty$ fractal.  Consequently, the accumulation locus ${\cal
  B}_y$ cannot cross the positive $y$ plane at any finite value of
$y$.

\begin{figure}[htbp]
\begin{center}
\includegraphics[height=6cm]{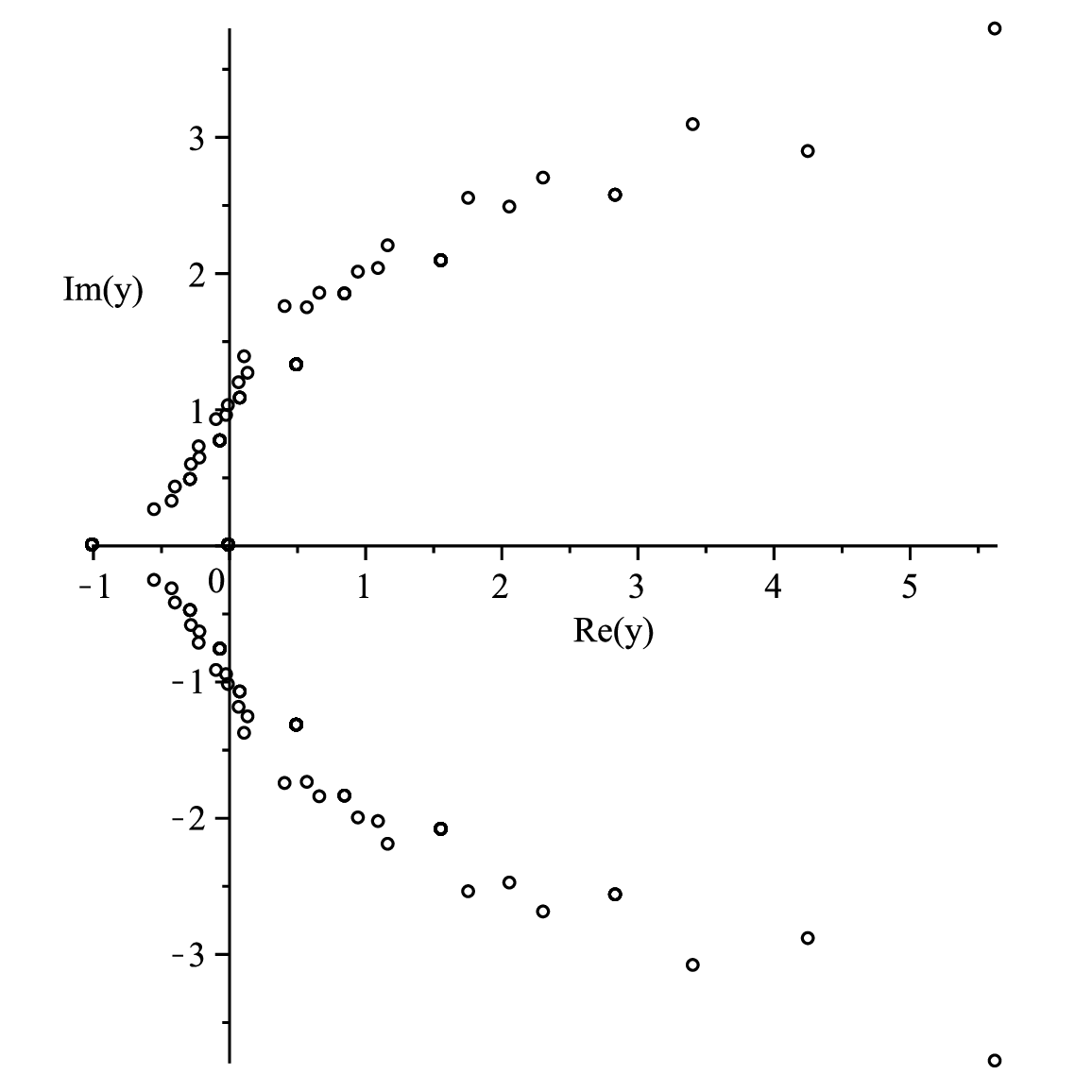}
\end{center}
\caption{\footnotesize{Zeros of $Z(H_4,q,y)$ in the $y$ plane for
$q=2$.}}
\label{hm4yplotq2_fig}
\end{figure}

\begin{figure}[htbp]
\begin{center}
\includegraphics[height=6cm]{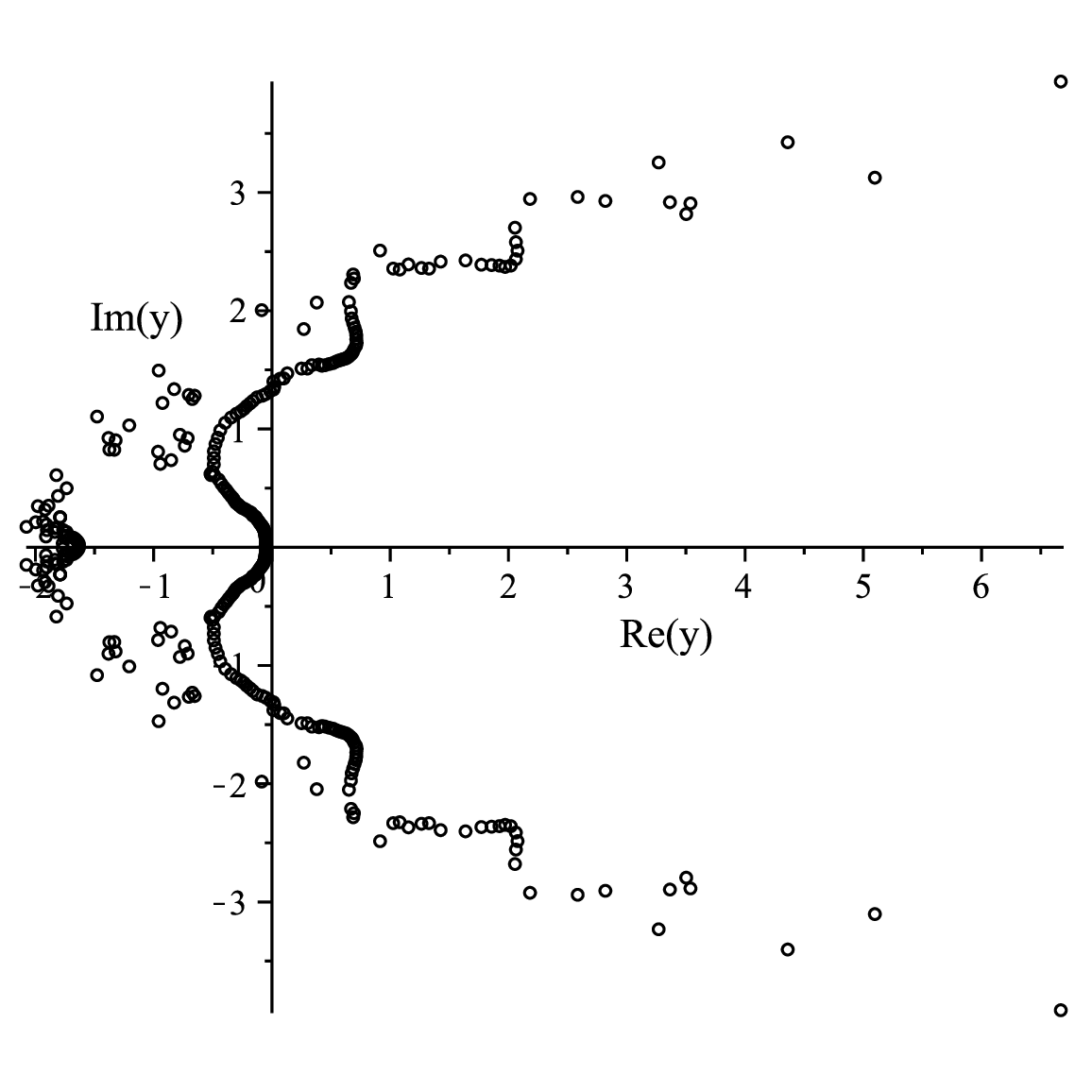}
\end{center}
\caption{\footnotesize{Zeros of $Z(H_4,q,y)$ in the $y$ plane for
$q=(1/2)(3+\sqrt{5} \, )$.}}
\label{hm4yplotqB5_fig}
\end{figure}

\begin{figure}[htbp]
\begin{center}
\includegraphics[height=6cm]{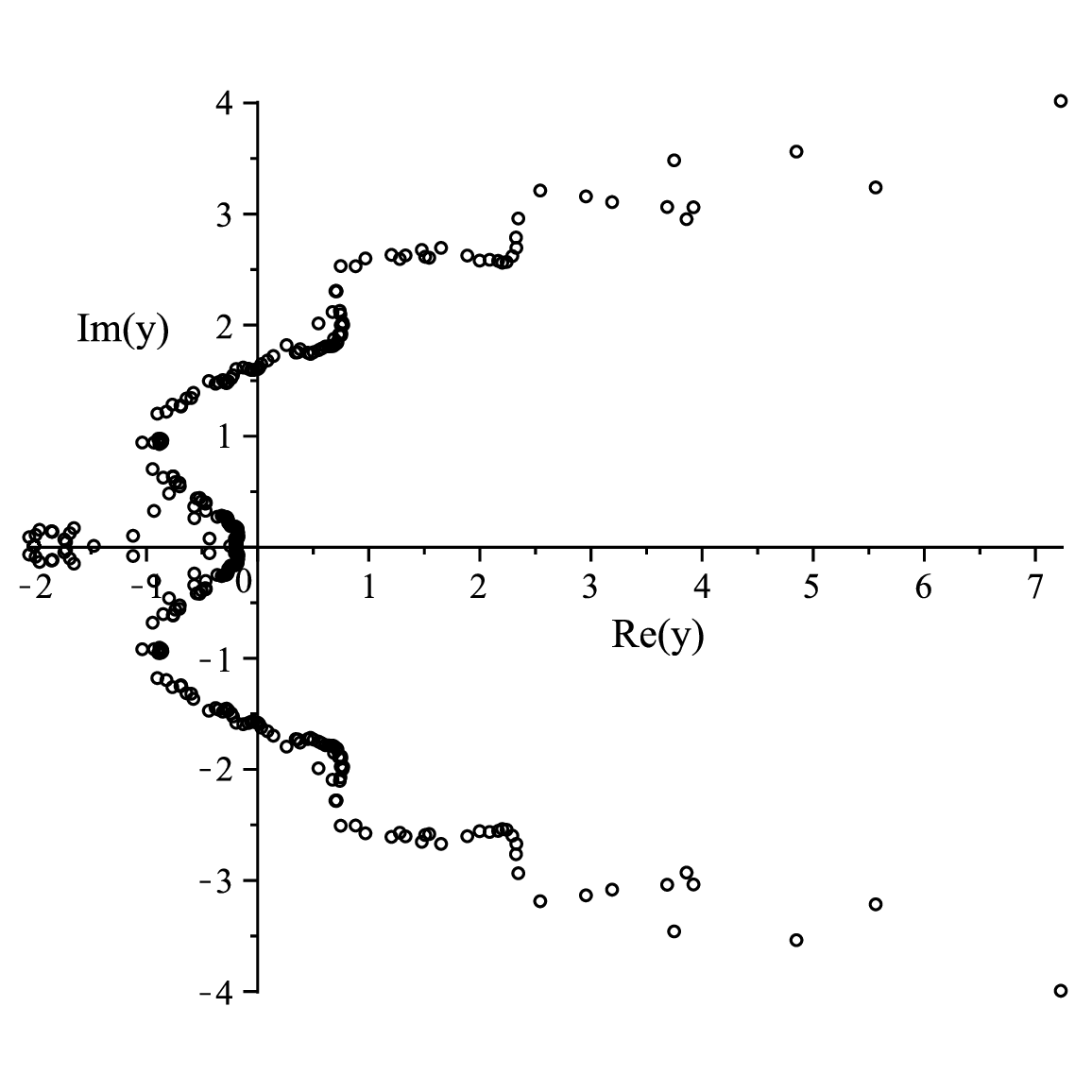}
\end{center}
\caption{\footnotesize{Zeros of $Z(H_4,q,y)$ in the $y$ plane for
$q=3$.}}
\label{hm4yplotq3_fig}
\end{figure}

\begin{figure}[htbp]
\begin{center}
\includegraphics[height=6cm]{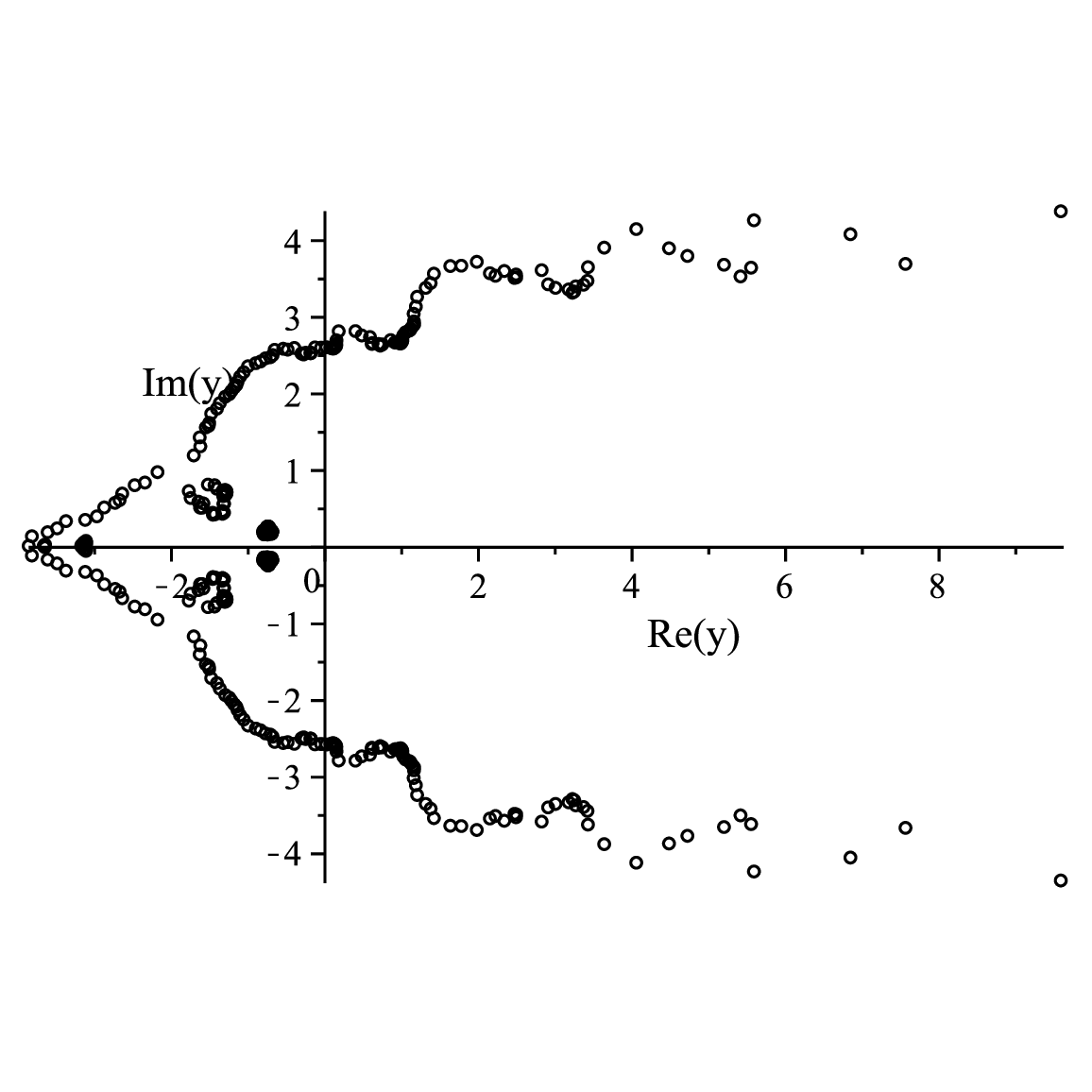}
\end{center}
\caption{\footnotesize{Zeros of $Z(H_4,q,v)$ in the $y$ plane for
$q=5$.}}
\label{hm4yplotq5_fig}
\end{figure}

In Figs. \ref{hm4yplotq2_fig}-\ref{hm4yplotq5_fig} we show zeros of
$Z(H_4,q,y)$ in the $y$ plane for $q=2, \ B_5, \ 3, \ 5$.  For each figure, the
number of zeros, counting multiplicity, is equal to the degree of $Z(H_m,q,y)$
in $y$, namely ${\rm deg}_y[Z(H_m,q,y)]=e(H_m)$, as given in
Eq. (\ref{e_Hm}). A remark is in order concerning the plot of zeros of
$Z(H_4,q=2,y)$ in Fig. \ref{hm4yplotq2_fig}.  For this $q=2$ case, we find that
$Z(H_m,q=2,y)$ has multiple zeros at $y=0$ and $y=-1$. For the cases that we
have calculated with $m \ge 1$, $Z(H_m,q=2,y)$ has a general form involving the
factor $y^{3^m}$ and, for $m \ge 1$, the factor $(y+1)^{(3/2)(3^m-1)}$, as well
as possible other repeated factors such as powers of $(y^2-y+2)$.  As
illustrative examples with $0 \le m \le 2$, we display
\beq 
Z(H_0,q=2,y)=2y(y^2+3)
\label{zhm0q2}
\eeq
\beq
Z(H_1,q=2,y) = 2y^3(y+1)^3(y^2-y+2)(y^4-2y^3+8y^2+2y+7)
\label{zhm1q2}
\eeq
and 
\beqs
&& Z(H_2,q=2,y) = 2y^9(y+1)^{12}(y^2-y+2)^3(y^4-3y^3+7y^2-y+4) \cr\cr
&\times& \Big (y^8-6y^7+26y^6-38y^5+84y^4-2y^3+102y^2+46y+43 \Big ) 
\label{zhm2q2}
\eeqs
The factor of $y^{3^m}$ in $Z(H_m,q=2,v)$ is the same as
for $Z(S_m,q=2,y)$ and reflects the fact that as $y \to 0$, i.e., $v \to -1$,
$Z(H_m,q,y) \to P(H_m,q)$, but $P(H_m,2)=0$ because it is not possible to
perform a proper vertex coloring of $H_m$ with just 2 colors.  Owing to these 
multiple zeros, the number of separate zeros in
Fig. \ref{hm4yplotq2_fig} is less than the total number of zeros, $e(H_4)=363$.
It is worthwhile observing here how, in the two-variable
polynomial $Z(H_m,q,y)$, one can see the approach to the zero at 
$Z(H_m,q=2,y=0)$ by setting $q=2$ and noting the presence of
the factor $y^{3^m}=(v+1)^{3^m}$ in $Z(H_m,q=2,y)$ or by setting 
$y=0$ and noting the presence of the factor $(q-2)^{3m}$ in
$Z(H_m,q,y=0)=P(H_m,q)$ (recall Eq. (\ref{phm_form})). 

If, for the Potts antiferromagnet, the asymptotic locus of chromatic
zeros ${\cal B}_q(y=0)$ crosses the real-$q$ axis at a maximal point
$q_c$, this connotes a zero-temperature critical point, so that the
corresponding asympotic locus of Fisher zeros, ${\cal B}_y$, should
cross the real $y$ axis at $y=0$.  Indeed, if one views this locus as
an algebraic variety in the ${\mathbb C}^2$ space of $(q,y)$, this
singular behavior occurs at a given point, $(q,y)=(q_c,0)$, and one is
observing ``slices'' through this algebraic variety with one or the
other variable held fixed in these crossings.  For example, the
property that $q_c=3$ for the infinite square lattice
\cite{lenard,lieb} and also for infinite-length, finite-width
square-lattice strips with self-dual boundary conditions
\cite{dg,sdg}, correspond to the $T=0$ critical point of the $q=3$
Potts antiferromagnet on these lattices. Similarly, the property that
$q_c(tri)=4$ for the triangular lattice \cite{baxter86,baxter87}
corresponds to the $T=0$ critical point of the PAF on this lattice,
and these results are in agreement with studies of Fisher zeros on
finite sections of these lattices, e.g., \cite{pfef,p,p2}. Among
fractals, $q_c=3$ for the Diamond Hierarchical Lattice $D_\infty$, and
this was shown to correspond to a $T=0$ critical point of the PAF on
$D_\infty$ \cite{dhl}. Thus, a consistency check on our inferred value
of $q_c(H_\infty)$ in Eq. (\ref{qc_hinf}) is to calculate Fisher zeros
of $Z(H_m,q,y)$ for $q=B_5$ and confirm that these are consistent with
the property that the asymptotic locus ${\cal B}_y$ passes through
$y=0$.  We have performed this check and show, as an example, the
zeros of $Z(H_4,q=B_5,y)$ in Fig. \ref{hm4yplotqB5_fig}.  As is
evident in this figure, even at the moderate iteration stage $m=4$,
these zeros pass very close to $y=0$, and are fully in accord with the
inference that as $m \to \infty$, the locus ${\cal B}_y$ for $q=B_5$
would pass through this point $y=0$. Parenthetically, we note that
these zeros in the $y$ plane for $q=B_5$ and $q=5$ (as well as larger
$q$, such as in Fig. \ref{hm4yplot_q10_fig} below) exhibit an
intriguing wavy structure as well as concentrations at (unphysical)
points and regions in the third and fourth quadrants.


\section{Large-$q$ Behavior of Zeros of $Z(H_m,q,y)$ in the $y$ Plane}
\label{lq_section}

One interesting result concerns the zeros in the large-$q$ limit. For regular
(non-fractal) lattices, these have previously been studied, e.g., in
\cite{wulq} and, by us, in \cite{lq}. In the thermodynamic limit for $d \ge 2$
on these regular lattices, the Potts ferromagnet has a finite-temperature phase
transition, so that ${\cal B}_y$ crosses the positive $y$ axis.  In contrast,
as was mentioned above, since the Potts model has no order-disorder phase
transition at any finite temperature on the $H_\infty$ fractal, the locus
${\cal B}_y$ does not cross the positive $y$ axis at any point. This is the
analogue, for Hanoi graphs, of the feature that we mentioned for Sierpinski
gasket graphs in \cite{sg}, that ${\cal B}_y$ cannot cross the positive $y$
axis for $S_\infty$.  We recall, however, that the Diamond Hierarchical Lattice
has infinite ramification number, as does the Sierpinski carpet
\cite{gam_ramified}, so that, as was discussed in \cite{ddi83} and more
recently in \cite{dhl}, ${\cal B}_y$ does cross the positive real $y$ axis for
the DHL fractal $D_\infty$.

\begin{figure}[htbp]
\begin{center}
\includegraphics[height=6cm]{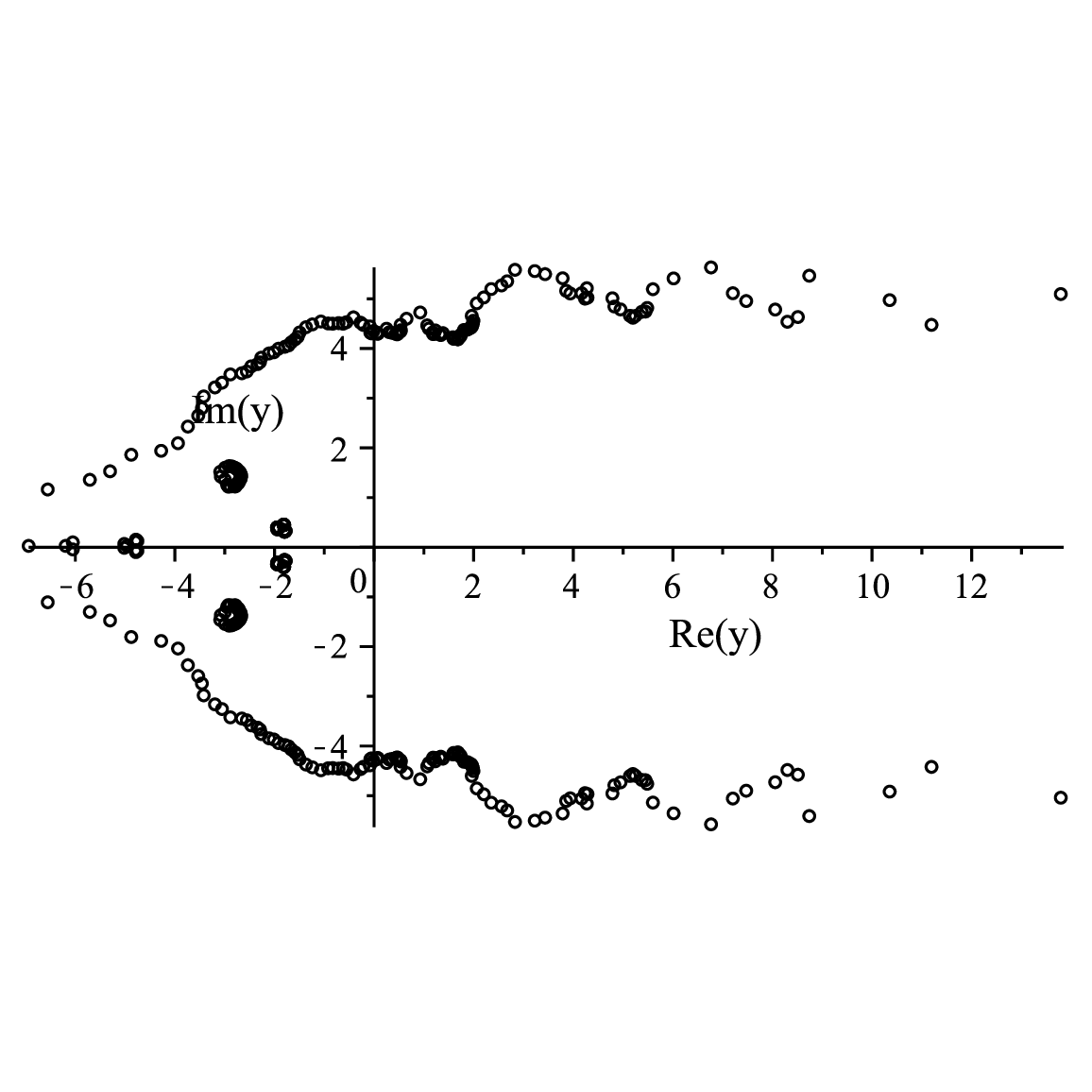}
\end{center}
\caption{\footnotesize{Zeros of $Z(H_4,q,y)$ in the $y$ plane for
$q=10$.}}
\label{hm4yplot_q10_fig}
\end{figure}

\begin{figure}[htbp]
\begin{center}
\includegraphics[height=6cm]{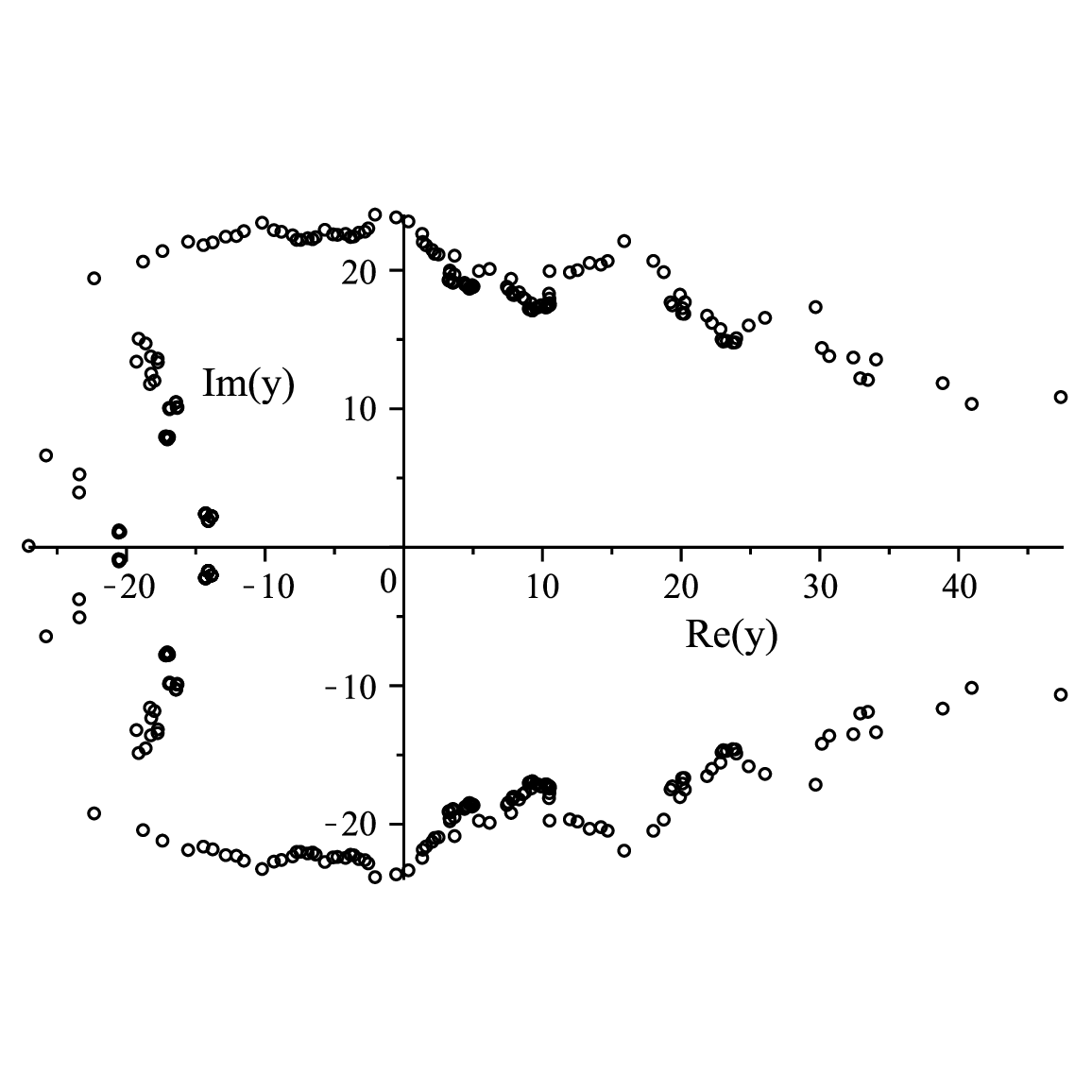}
\end{center}
\caption{\footnotesize{Zeros of $Z(H_4,q,y)$ in the $y$ plane for
$q=10^2$.}}
\label{hm4yplot_q100_fig}
\end{figure}

\begin{figure}[htbp]
\begin{center}
\includegraphics[height=6cm]{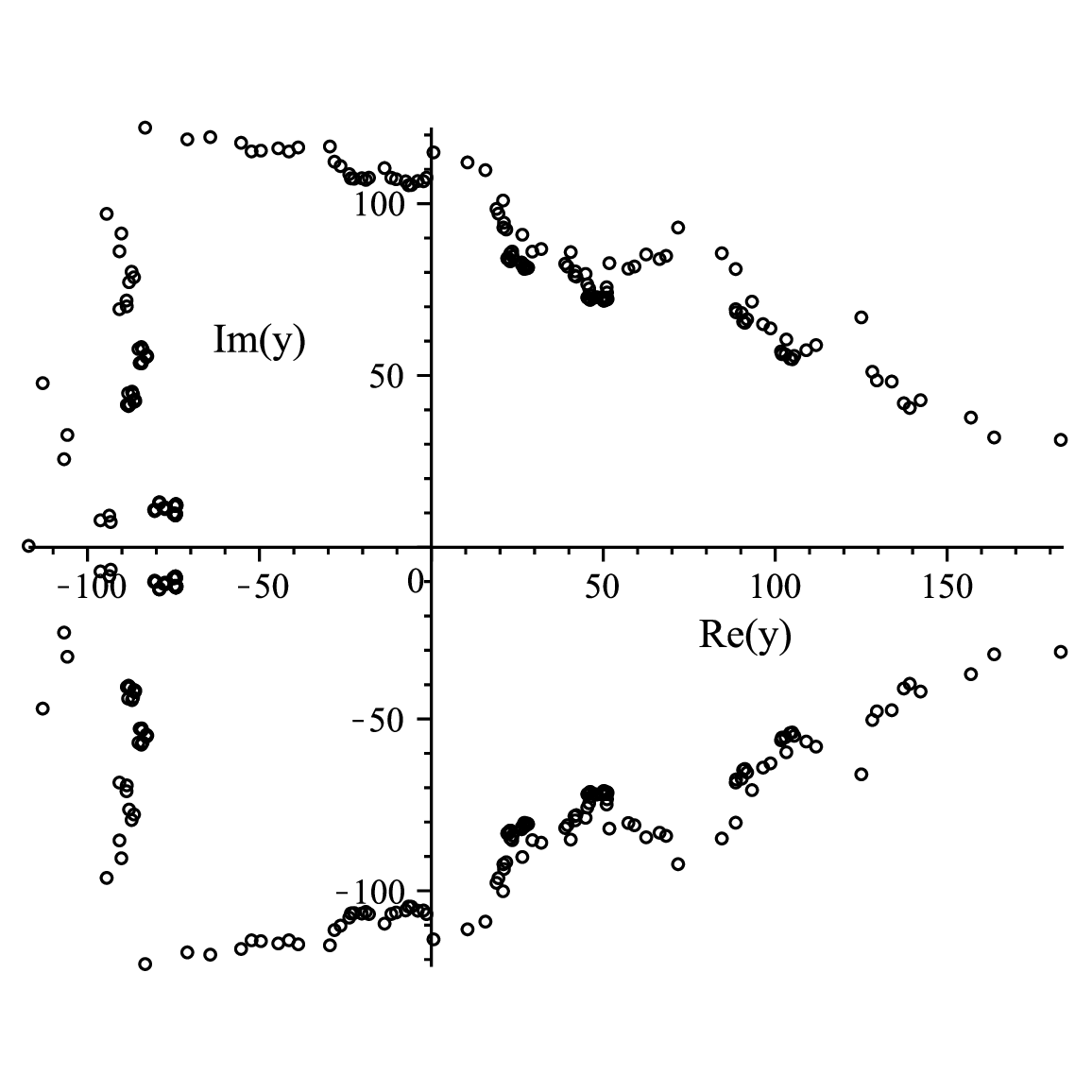}
\end{center}
\caption{\footnotesize{Zeros of $Z(H_4,q,y)$ in the $y$ plane for
$q=10^3$.}}
\label{hm4yplot_q1000_fig}
\end{figure}

\begin{figure}[htbp]
\begin{center}
\includegraphics[height=6cm]{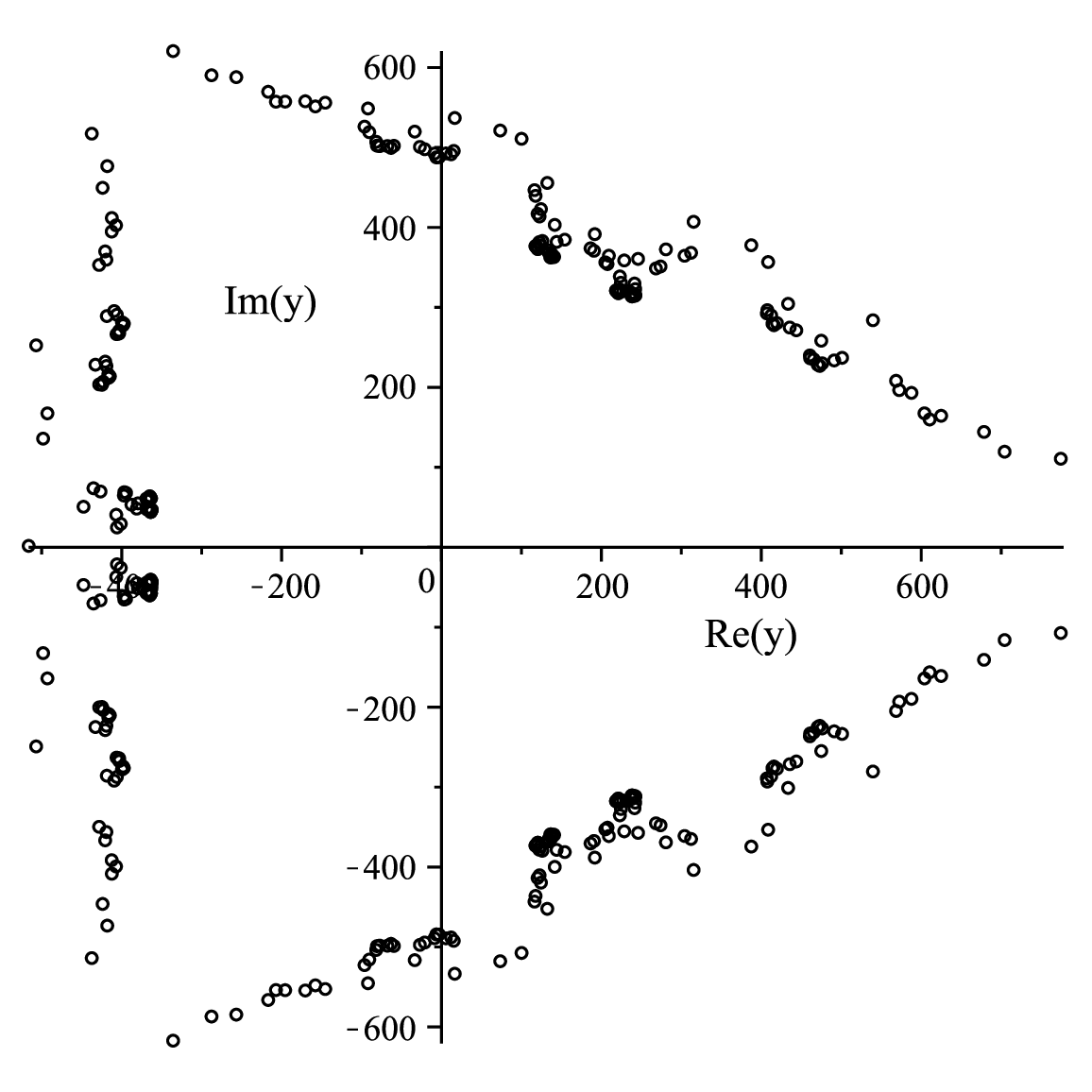}
\end{center}
\caption{\footnotesize{Zeros of $Z(H_4,q,y)$ in the $y$ plane for
$q=10^4$.}}
\label{hm4yplot_q10000_fig}
\end{figure}

\begin{figure}[htbp]
\begin{center}
\includegraphics[height=6cm]{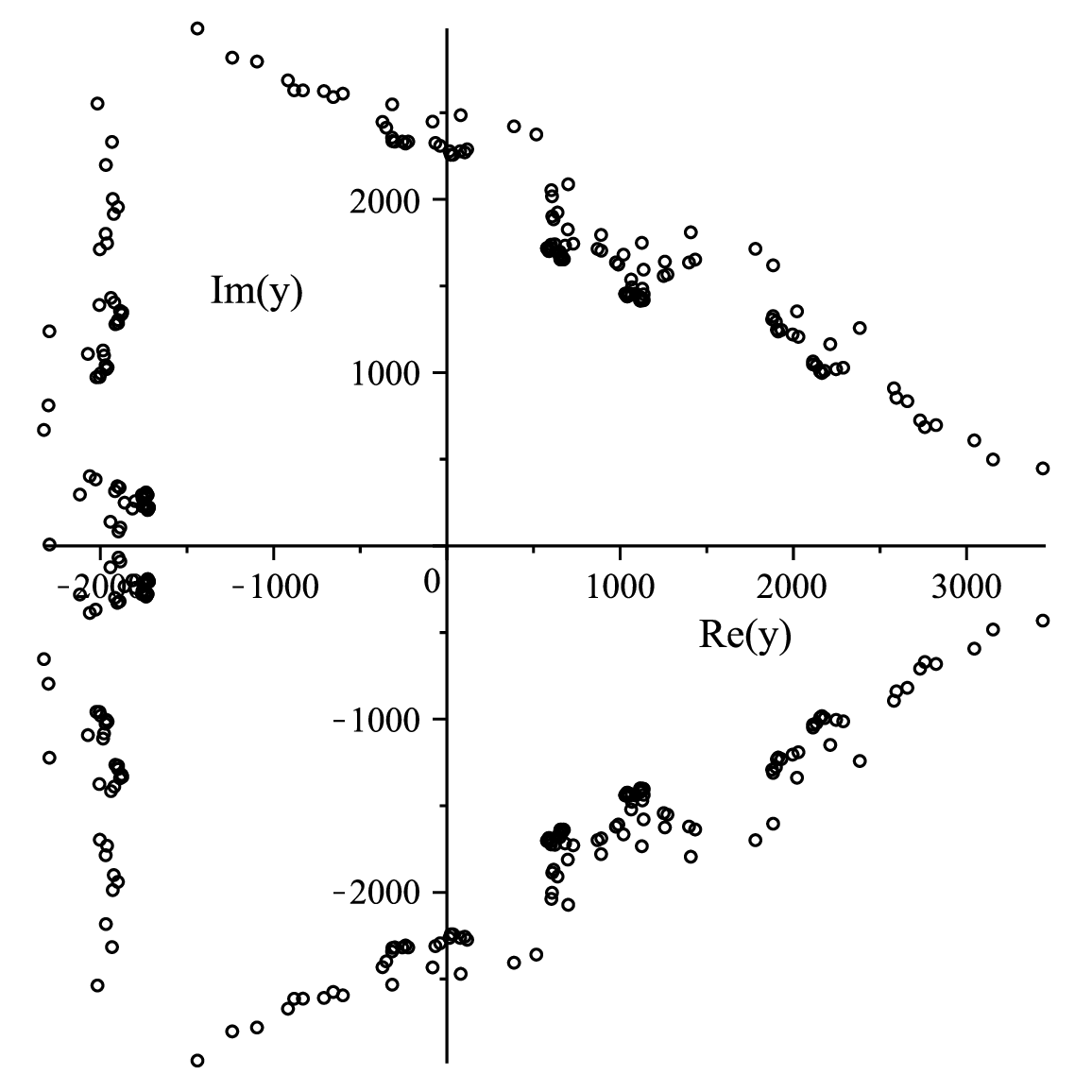}
\end{center}
\caption{\footnotesize{Zeros of $Z(H_4,q,y)$ in the $y$ plane for
$q=10^5$.}}
\label{hm4yplot_q100000_fig}
\end{figure}

In Figs. \ref{hm4yplot_q10_fig}-\ref{hm4yplot_q100000_fig} we show plots of
zeros of $Z(H_4,q,y)$ in the $y$ plane for $q=10$, $10^2$, $10^3$, $10^4$, and
$10^5$.  We find that, in contrast with the sections of regular lattices that
were studied in Refs. \cite{wulq,lq}, where the $y$-plane zeros in the
large-$q$ limit approach an approximately circular form with $|y| \simeq
q^{2/\Delta}$, here we find a different type of behavior, namely that, for $q
\gg 1$, the zeros cluster approximately along, or near to, parts of the edges
of an equilateral triangle with vertices at points that scale like $y_j \sim
q^{2/3} e^{\frac{2\pi ij}{3}}$, where $j=0, \ 1, \ 2$, or equivalently, 
\beq
v_1 \sim q^{2/3}, \quad v_{2,3} \sim q^{2/3}e^{\pm \frac{2 \pi i}{3}} \ .
\label{vjlq}
\eeq
These are equivalent because the magnitudes of these zeros grow like
$|q|^{2/3}$ in the large-$q$ limit, and hence there is a negligibly small
difference between the positions of the zeros in the $y$ plane and the $v=y-1$
plane.  However, the zeros avoid the regions around the three apex points of
this triangle; in particular, they avoid the apex point on the positive $y$
axis, as noted above.

\begin{figure}[htbp]
\begin{center}
\includegraphics[height=6cm]{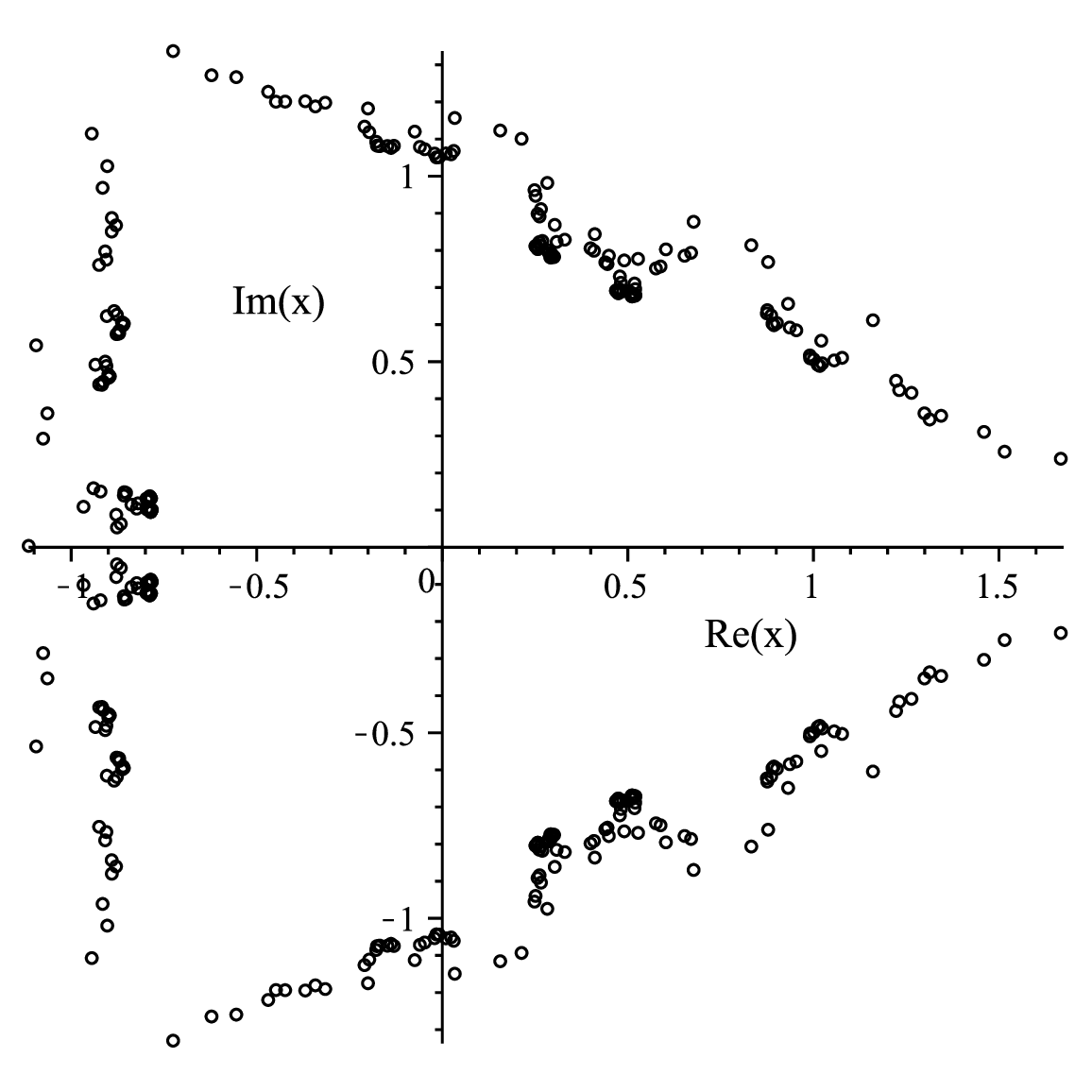}
\end{center}
\caption{\footnotesize{Zeros of $Z(H_4,q,y)$ in the $\xi=v/q^{2/3}$ plane for
$q=10^4$.}}
\label{hm4xplot_q10000_fig}
\end{figure}

\begin{figure}[htbp]
\begin{center}
\includegraphics[height=6cm]{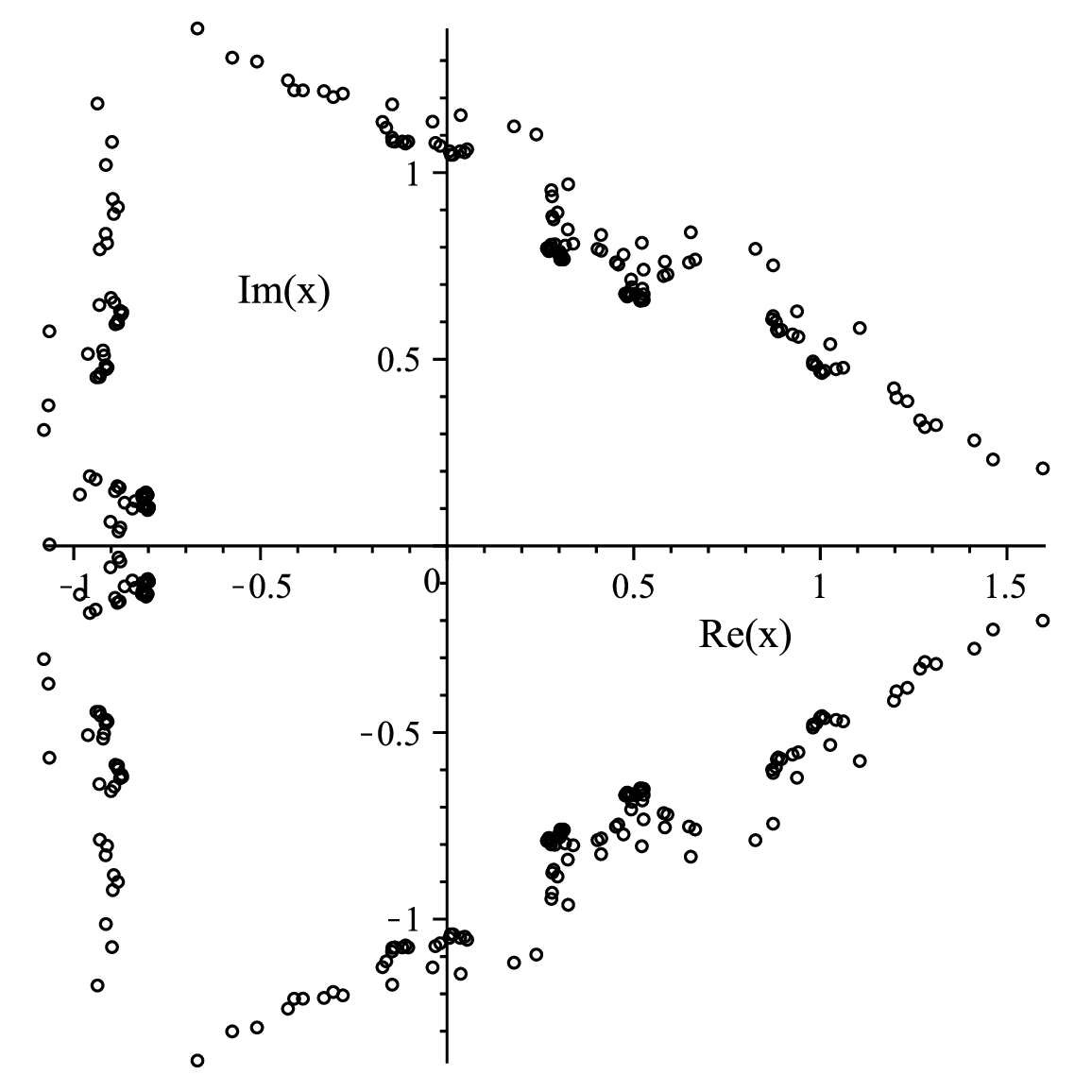}
\end{center}
\caption{\footnotesize{Zeros of $Z(H_4,q,y)$ in the $\xi=v/q^{2/3}$ plane for
$q=10^5$.}}
\label{hm4xplot_q100000_fig}
\end{figure}

It is also instructive to display these zeros as a function of a variable
\beq
\xi = \frac{v}{q^{2/3}} \ , 
\label{xi}
\eeq
which, up to negligibly small terms in the large-$q$ limit, is equivalent to 
$\xi = y/q^{2/3}$. We show these plots for $q=10^4$ and $q=10^5$ in Figs.
\ref{hm4xplot_q10000_fig} and \ref{hm4xplot_q100000_fig}.  The apex points in 
this $\xi$ plane have magnitudes $|\xi| \simeq 2$. An important property of
these zeros is invariance under the action of a multiplicative 
${\mathbb Z}_3$ group with the elements
\beq
{\mathbb Z}_3: \quad \{1, \ e^{2\pi i/3}, \ e^{4\pi i/3} \} \ , 
\label{z3group}
\eeq
including rotations in the complex $\xi$ plane by an angle of $\pm 2\pi/3$ 
radians. 

We provide some insight into this behavior as follows.  We will show that in
the limit $q \to \infty$, the zeros of $Z(H_m,q,v)$ are determined by a
polynomial in the variable
\beq
\eta = \xi^3 = \frac{v^3}{q^2} \ . 
\label{eta}
\eeq
This also explains why, for a given large $q$, the zeros in the $v$ plane or
equivalently, the $\xi$ plane, are invariant under the action of the elements
of the ${\mathbb Z}_3$ group (\ref{z3group}). Consider, for example,
$Z(H_1,q,v)$, the expression for which is given in Eq. (\ref{zhm1}) in Appendix
\ref{tcal_appendix}. Expressing this in terms of $\xi$ and expanding as 
$q \to \infty$, we get 
\beqs
Z(H_1,q,v) &=& q^9\Bigg [ \Big ( \xi^{12} + \xi^9 + 3\xi^6 + 3\xi^3 + 1 \Big ) 
\cr\cr
&+& \frac{3}{q^{1/3}}\Big ( 2\xi^{10} + 6\xi^7 + 9\xi^4 + 4\xi \Big ) \cr\cr
&+& \frac{6}{q^{2/3}}\Big ( 2\xi^{11} + 8\xi^8 +18\xi^5 + 11\xi^2 \Big ) \cr\cr
&+& \frac{1}{q}\Big ( 75\xi^9 + 247\xi^6 + 217\xi^3 \Big ) + 
O\Big ( \frac{1}{q^{4/3}}\Big ) \quad \Bigg ] \ , 
\label{zhm1xi} 
\eeqs
where the additional terms are polynomials in $\xi$ multiplied by negative 
powers $q^{-4/3}$, $q^{-5/3}$. $q^{-2}$, etc. Recalling Eq. (\ref{n_Hm}),
this example shows in the limit $q \to \infty$,
\beq
\lim_{q \to \infty} \frac{Z(H_1,q,v)}{q^{n(H_1)}} = \Omega(H_1,\eta)
\label{zhm1xi_qinf}
\eeq
where 
\beq
\Omega(H_1,\eta) = \eta^4 + \eta^3 + 3\eta^2 + 3\eta + 1 \ . 
\label{Omega_Hm1}
\eeq
This method generalizes to higher $m$, which shows that 
\beq
\lim_{q \to \infty} \frac{Z(H_m,q,v)}{q^{n(H_m)}} = \Omega(H_m,\eta)
\label{zhmxi_qinf}
\eeq
where $n(H_m)=3^{m+1}$ was given in Eq. (\ref{n_Hm}) and $\Omega(H_m,\eta)$ is
a polynomial in $\eta$.  For example, for the next higher iterate, $H_2$, 
we calculate 
\beqs
&&\Omega(H_2) = (\eta+1)(\eta^{12}+3\eta^{10}+9\eta^9+19\eta^8+38\eta^7
+58\eta^6 \cr\cr
&+&71\eta^5+73\eta^4+56\eta^3+28\eta^2+8\eta+1) \ .
\label{Omega_Hm2}
\eeqs
Because, in the limit $q \to \infty$, $Z(H_m,q,v)$
reduces to the prefactor $q^{n(H_m)}$ times a function of $\xi^3$ and thus of
$v^3$, this shows that the zeros of $Z(H_m,q,v)$ in this limit are invariant
under the elements of the ${\mathbb Z}_3$ group (\ref{z3group}) in the $\xi$
plane, and equivalently, in the $v$ and $y$ planes. The property that these
zeros accumulate approximately along parts of the edges of the equilateral
triangle with apex points (\ref{vjlq}) depends on further details of
$Z(H_m,q,v)$.

It is useful to contrast these findings with the results that we obtained in
Ref. \cite{lq}.  In that work we discussed the analytic basis that is
responsible for the approach of the Potts partition function zeros in the
complex plane of the variable $\xi = v/q^{2/\Delta}$ to the circle $|\xi|=1$,
as $q \to \infty$, for regular lattices with vertex degree $\Delta$ (or
$\Delta_{\rm eff}$). This circular locus of the accumulation set of zeros in
the $\xi$ plane is invariant under the full rotation group 
U(1) $\approx$ O(2). We also showed this behavior for the Sierpinski iterates
in \cite{sg} and for the DHL iterates in \cite{dhl} with $\Delta_{\rm eff}$. In
addition to our derivation given above, an explanation for why the zeros do not
cluster on or near to this circle is provided by our analysis in \cite{lq}.
There we noted that our derivation only applied to lattice graphs with the
property that deleting several edges would not lead to the appearance of
disconnected graphical components.  As stated in \cite{lq}, this condition is
satisfied for sufficiently large sections of lattice graphs with finite 
aspect ratios $L_i/L_j$, where $L_i$ and $L_j$ denote lengths along two
different lattice directions. As we
remarked in \cite{sg,dhl}, this condition is also true for the Sierpinski
gasket iterates $S_m$ and the DHL iterates $D_m$. However, in contrast, it is
not true for the Hanoi iterates $H_m$; if one deletes any two of the edges on
the middle exterior sides of the outer triangular boundary, this separates the
previously connected graph $H_m$ into two disjoint components. Our detailed
analysis above shows that the symmetry of the zeros in the $\xi$ plane in the
$q \to \infty$ limit for $H_m$ graphs is the finite subgroup ${\mathbb Z}_3$ of
the full rotation group U(1). 


\section{Conclusions}
\label{conclusion_section}

In summary, in this work we have investigated properties of the Potts model
partition function $Z(H_m,q,v)$ on $m$'th iterate Hanoi graphs and have used
the results to draw inferences about the $m \to \infty$ limit.  We have
calculated the ground state degeneracy per vertex of the Potts antiferromagnet
on $H_m$ for $q=3$ and $q=4$ and a large range of $m$ and have used the results
to infer estimates of $W(H_\infty,q)$ for these values of $q$. The values were
compared with the corresponding ground state degeneracy for the Potts
antiferromagnet on other lattices. Further, we have presented calculations of
zeros of $P(H_m,q)$ for $m$ up to 4, and from these we have inferred that in
the $m \to \infty$ limit, the asymptotic accumulation locus of chromatic zeros,
${\cal B}_q$ for $v=y-1=-1$, crosses the real $q$ axis at $q_c(H_\infty) = B_5
= (1/2)(3+\sqrt{5} \, )$.  This means that the Potts antiferromagnet with
$q=B_5$ has a zero-temperature critical point on the $H_\infty$ fractal. We
have obtained further evidence in support of this inference by calculating the
partition function zeros in the $y$ plane for this value of $q$ and showing
that they are consistent with the inference that the locus of zeros in the
limit $m \to \infty$ passes through the $T=0$ point for the antiferromagnetic
Potts model, at $y=0$.  Results were also given for the zeros of the partition
function on $H_m$ (i) in the $q$ plane for the Potts antiferromagnet and
ferromagnet at illustrative finite temperatures and (ii) in the $y$ plane for
several values of $q$ in addition to $B_5$.  Finally, we have computed the
zeros in the $y$ plane for $q \gg 1$ and have shown that they aggregate
approximately along parts of the sides of the triangle whose apex points scale
like $q^{2/3}$ and $q^{2/3}e^{\pm 2 \pi i/3}$, exhibiting an invariance under
elements of the multiplicative ${\mathbb Z}_3$ group.  Some comparisons were
made with our earlier work on the related Sierpinski fractal.


\begin{acknowledgments}

The research of S.-C.C. was supported in part by the Taiwan Ministry of
Science and Technology (MOST) grant MOST 111-2115-M-006-012-MY2 and the
Taiwan National Science and Technology Council (NSTC) grant NSTC
113-2115-M-006-006-MY2. The research of
R.S. was supported in part by the U.S. National Science Foundation Grant
NSF-22-10533.

\end{acknowledgments}


\begin{appendix}


\section{Iterative Procedure for Calculation of $T(H_m,x,y)$}
\label{tcal_appendix}

For reference, here we remark on the nonlinear iterative procedure
derived in \cite{donno_iacono} for calculating the Tutte polynomial
$T(H_m,x,y)$. (Ref. \cite{donno_iacono} used a different labelling convention
for the Hanoi graphs, according to which $H_{m'}$ in \cite{donno_iacono} is 
$H_{m+1}$ in the labelling convention used here, so $m'=m+1$.) 
This procedure expresses $T(H_m,x,y)$ in terms of a sum
of three auxiliary functions $F_{0,m}(x,y)$, $F_{1,m}(x,y)$, and 
  $F_{2,m}(x,y)$, which satisfy nonlinear recursive relations with
lower-order auxiliary functions. Explicitly (suppressing the arguments
in $F_{s,m} \equiv F_{s,m}(x,y)$, $s=0,1,2$),
\beq
T(H_m,x,y)=F_{2,m}+3F_{1,m}+F_{0,m}
\label{thmrel}
\eeq
where
\beqs
F_{2,m+1} &=& (y-1)F_{2,m}^3 + 3(x-1)^{-1}F_{2,m}F_{1,m}
\Big (2F_{2,m}+F_{1,m}\Big )
\cr\cr
&+& 3F_{2,m}(F_{2,m}+F_{1,m})^2
\label{F2_recursion}
\eeqs
and
\beqs
&& F_{1,m+1} = (y-1)F_{2,m}^2F_{1,m} \cr\cr
&+&(x-1)^{-1}\Big (F_{2,m}^2F_{0,m} + 7F_{2,m}F_{1,m}^2+2F_{2,m}F_{1,m}F_{0,m}
+4F_{1,m}^3+F_{1,m}^2F_{0,m} \Big ) \cr\cr
&+& 7F_{2,m}^2F_{1,m}+2F_{2,m}^2F_{0,m}+14F_{2,m}F_{1,m}^2
+4F_{2,m}F_{1,m}F_{0,m}+7F_{1,m}^3+2F_{1,m}^2F_{0,m} \cr\cr
&+&(x-1)\Big ( F_{2,m}^3+5F_{2,m}^2F_{1,m}+F_{2,m}^2F_{0,m}
+7F_{2,m}F_{1,m}^2+2F_{2,m}F_{1,m}F_{0,m} \cr\cr
&+&3F_{1,m}^3+F_{1,m}^2F_{0,m} \Big ) 
\label{F1_recursion}
\eeqs
with the initial values
\beq
F_{2,0}=y+2, \quad F_{1,0}=x-1, \quad F_{0,0}=(x-1)^2 \ .
\label{initial}
\eeq
Since these expressions are nonsingular at $x=1$, the presence of the
factors $(x-1)^{-1}$ implies certain identities. For example, in
Eq. (\ref{F2_recursion}), the quantity
$F_{2,m}F_{1,m}(2F_{2,m}+F_{1,m})=0$ at $x=1$, and similarly with the
quantity multiplying the factor of $(x-1)^{-1}$ in Eq.
(\ref{F1_recursion}).  The recursion relation for $F_{0,m+1}$ is
longer, and we refer the reader to (Theorem 4.3 of)
Ref. \cite{donno_iacono} for it.  From the Tutte polynomial one can
calculate the Potts model partition function via the relation
(\ref{ztrel}) with (\ref{xqv}) and (\ref{yqv}).
For our work we have calculated $T(H_m,x,y)$ with $m$ to 4 inclusive. 
As an illustration, we list $T(H_1,x,y)$:
\beqs
&& T(H_1,x,y) = x^8+4x^7+3x^6y+7x^6+9x^5y+3x^4y^2+8x^5+12x^4y \cr\cr
&+&6x^3y^2+x^2y^3+8x^4+13x^3y+9x^2y^2+4xy^3+y^4+7x^3+12x^2y \cr\cr
&+&9xy^2+3y^3+4x^2+6xy+3y^2+x+y \ .
\label{thm1}
\eeqs
The corresponding expression for $Z(H_1,q,v)$ is
\beqs
&&Z(H_1,q,v) = q\bigg [ 
v^{12}+12v^{11}+6v^{10}q+v^9q^2+60v^{10}+75v^9q+48v^8q^2 \cr\cr
&+&18v^7q^3+3v^6q^4+144v^9+312v^8q+351v^7q^2+247v^6q^3 \cr\cr
&+&108v^5q^4+27v^4q^5+3v^3q^6+135v^8+423v^7q+674v^6q^2 \cr\cr
&+&684v^5q^3+468v^4q^4+217v^3q^5+66v^2q^6+12vq^7+q^8 \  \bigg ] \ . 
\label{zhm1}
\eeqs
The expressions for $T(H_m,x,y)$ and $Z(H_m,q,v)$ become quite lengthy for
higher $m$, so we do not list these explicitly here. For example, while
$T(H_1,x,y)$ and $Z(H_1,q,v)$ have 24 and 25 terms, respectively, as displayed
in Eqs. (\ref{thm1}) and (\ref{zhm1}), $T(H_2,x,y)$ and $Z(H_2,q,v)$ have 195
and 196 terms, respectively, and so forth for higher $m$.


\section{Sierpinski Gasket Graphs}
\label{sg_appendix}

In this Appendix we list some properties of $m$'th iterates of Sierpinski
gasket graphs, $S_m$.  Our labelling convention for the $S_m$ is the same as we
used in \cite{sg} and \cite{sts}-\cite{iss}. (This also maintains uniformity
with our labelling convention for the $H_m$, so that both $H_0$ and $S_0$ are
the same graph, namely $K_3=C_3$.)

The number of vertices in the $m$'th iterate Sierpinski graph, $n(S_m)$, is
\beq
n(S_m) = \frac{3(3^m+1)}{2} \ ,  
\label{n_Sm}
\eeq
and the numbers of edges, $e(S_m)$, is 
\beq
e(S_m) = 3^{m+1} \ .
\label{e_Sm}  
\eeq
In comparison with the Hanoi iterates, we thus have
\beq
\lim_{m \to \infty} \, \frac{n(H_m)}{n(S_m)} = 2
\label{n_ratio_hs}
\eeq
and
\beq
\lim_{m \to \infty} \, \frac{e(H_m)}{e(S_m)} = \frac{3}{2} \ .
\label{e_ratio_hs}
\eeq

The cyclomatic number of $S_m$ is thus 
\beq
c(S_m) = \frac{3^{m+1}-1}{2} \ .
\label{c_Sm}
\eeq
The effective vertex degree of $S_\infty$ is
\beq
\Delta_{\rm eff}(S_\infty) = 4 \ .
\label{delta_Sinf}
\eeq
The number of faces in $S_m$, denoted $N_F(S_m)$, is 
\beq
N_F(S_m) = \frac{3^{m+1}-1}{2} \ ,
\label{faces_Sm}
\eeq
which is equal to $c(S_m)$.  The number of triangles in $S_m$, denoted
$N_t(S_m)$, is given by the coefficient of the term
$z^m$ in the Taylor series expansion of the function $(1+z)/(1-3z)$
about $z=0$, i.e.,
\beqs
\frac{1+z}{1-3z} &=& \sum_{j=0}^\infty N_t(S_j) z^j = 
1 + 4z + 12z^2 + 36z^3 + 108z^4 \cr\cr
&+& 324z^5 + 972z^6 + 2916z^7 + 8748z^8 + ...
\label{triangles_Sm}
\eeqs
Therefore, in the limit $m \to \infty$, the ratio of triangular faces to the
total number of faces in $S_m$ is
\beq
\lim_{m \to \infty} \frac{N_t(S_m)}{N_F(S_m)} = \frac{8}{9} \ .
\label{x}
\eeq
This is evidently slightly higher than the value of 2/3 for the corresponding
ratio for Hanoi graphs.

The Tutte polynomial for $m=0$ is elementary: $T(S_0,x,y)=T(K_3,x,y)=
x^2+x+y$. For $m=1$,
\beqs
T(S_1,x,y) &=& x^5+4x^4+4x^3y+3x^2y^2+3xy^3+y^4+6x^3+9x^2y+6xy^2 \cr\cr
           &+&2y^3+4x^2+6xy+3y^2+x+y
\label{tsgm1}
\eeqs
This yields
\beq
P(S_1,q)=q(q-1)(q-2)^4 \ .
\label{psgm1}
\eeq
As we calculated for the work in \cite{sg},
\beqs
P(S_2,q) &=& q(q-1)(q-2)^6\Big (q^7-14q^6+85q^5-292q^4+620q^3 \cr\cr
&-&831q^2+676q-272 \Big ) \ .
\label{psgm2}
\eeqs
and
\beqs
&&P(S_3,q) = q(q-1)(q-2)^9\bigg ( q^{31}-62q^{30}+1864q^{29}-36200q^{28} \cr\cr
&+&510406q^{27}-5567417q^{26}+48885472q^{25}-354996791q^{24} \cr\cr
&+&2173710199q^{23}-11385918177q^{22}+51580729311q^{21} \cr\cr
&-&203815546118q^{20}+707080076667q^{19}-2164599135972q^{18} \cr\cr
&+&5869749718724q^{17}-14137926421037q^{16}+30300472680589q^{15} \cr\cr
&-&57835231423884q^{14}+98313224299548q^{13}-148699316658336q^{12} \cr\cr
&+&199737162065052q^{11}-237551626238256q^{10}+249080624015424q^9 \cr\cr
&-&228900210474672q^8+182904284767200q^7-125717659569984q^6 \cr\cr
&+&73230886710720q^5-35383634429696q^4+13722399529984q^3 \cr\cr
&-&4041086324736q^2+811162107904q-84017414144 \bigg ) \ . 
\label{psgm3}
\eeqs
In \cite{sg} we calculated $P(S_m,q)$ for higher $m$, but the expressions
were too lengthy to give there. 

With the notation of Eq. (\ref{zvy}), for the cases that we calculated, we find
that $Z(S_m,2,y)$ has the factors $y^{3^m}$ and, for $m \ge 1$,
$(y^2+1)^{3^{m-1}}$, where $y=v+1$.  The factor of $y^{3^m}$ is the same as for
$Z(H_m,2,y)$ and reflects the property that as $y \to 0$, $Z(S_m,q,y) \to
P(S_m,q)$, but $P(S_m,2)=0$ because it is not possible to perform a proper
vertex coloring of $S_m$ with just 2 colors. As illustrations of the explicit
expressions of $Z(S_m,2,y)$ for the first few values of $m$, we 
note that $Z(S_0,2,y)=Z(H_0,2,y)$, given in Eq. (\ref{zhm0q2}), and list the
following:
\beq
Z(S_1,2,y)= 2y^3(y^2+1)(y^4+2y^2+13)
\label{zsgm1q2}
\eeq
\beq
Z(S_2,2,y) = 2y^9(y^2+1)^3(y^4+7)(y^8+26y^4+72y^2+157)
\label{zsgm2q2}
\eeq
and
\beqs
&&Z(S_3,2,y)=2y^{27}(y^2+1)^9(y^4+7)^3 ( y^8-2y^6+16y^4+34y^2+79 )
\cr\cr
&\times&
\Big ( y^{16}-4y^{14}+48y^{12}+124y^{10}+1034y^8+3988y^6+12696y^4+23156y^2+24493
\Big ) \ .
\cr\cr
&&
\label{zsgm3q2}
\eeqs

\end{appendix}




\end{document}